\documentclass[9pt,twocolumn,twoside]{paper}

\templatetype{researchpaper} 

\title{Cut it out: Out-of-plane stresses in cell sheet folding of \emph{Volvox} embryos}

\author[a,b,c,1,2]{Pierre A. Haas} 
\author[d,1,2]{Stephanie S. M. H. Höhn}

\affil[a]{Max Planck Institute for the Physics of Complex Systems, Nöthnitzer Stra\ss e 38, 01187 Dresden, Germany}
\affil[b]{Max Planck Institute of Molecular Cell Biology and Genetics, Pfotenhauerstra\ss e 108, 01307 Dresden, Germany}
\affil[c]{Center for Systems Biology Dresden, Pfotenhauerstra\ss e 108, 01307 Dresden, Germany}
\affil[d]{Department of Applied Mathematics and Theoretical Physics, University of Cambridge, Wilberforce Road, Cambridge CB3 0WA, United Kingdom}

\leadauthor{Haas \& Höhn} 

\significancestatement{During the development of multicellular organisms, cell shape changes drive the emergence of complex shapes by generating out-of-plane stresses that cause tissues to fold, but quantifying such out-of-plane stresses in curved tissues has remained challenging. Here, we develop a framework combining microsurgery experiments and a mechanical model to quantify these stresses and infer mechanical properties of the tissue based on the unfurling of the tissue on being cut. We find such recoils in the alga \emph{Volvox} and use our model to predict the stresses required to reproduce the observed tissue shape sequence quantitatively, both before and after cutting. Our results emphasise the effects of even small out-of-plane tissue stresses in development.}

\authorcontributions{S.M.H.H. and P.A.H. designed the study. S.M.H.H. performed experiments. P.A.H. derived the theoretical models. S.M.H.H. and P.A.H. analysed and interpreted the data and wrote the manuscript.}
\authordeclaration{The authors declare that they have no competing interests.}
\equalauthors{\textsuperscript{1}P.A.H. and S.M.H.H. contributed equally.}
\correspondingauthor{\textsuperscript{2}Joint corresponding authors. E-mails: haas@pks.mpg.de, s.hoehn@damtp.cam.ac.uk.}


\begin{abstract}
The folding of cellular monolayers pervades embryonic development and disease. It results from stresses out of the plane of the tissue, often caused by cell shape changes including cell wedging via apical constriction. These local cellular changes need not however be compatible with the global shape of the tissue. Such geometric incompatibilities lead to residual stresses that have out-of-plane components in curved tissues, but the mechanics and function of these out-of-plane stresses are poorly understood, perhaps because their quantification has proved challenging. Here, we overcome this difficulty by combining laser ablation experiments and a mechanical model to reveal that such out-of-plane residual stresses exist and also persist during the inversion of the spherical embryos of the green alga \emph{Volvox}. We show how to quantify the mechanical properties of the curved tissue from its unfurling on ablation, and reproduce the tissue shape sequence at different developmental timepoints quantitatively by our mechanical model. Strikingly, this reveals not only clear mechanical signatures of out-of-plane stresses associated with cell shape changes away from those regions where cell wedging bends the tissue, but also indicates an adaptive response of the tissue to these stresses. Our results thus suggest that cell sheet folding is guided mechanically not only by cell wedging, but also by out-of-plane stresses from these additional cell shape changes.
\end{abstract}

\dates{This manuscript was compiled on \today.}

\newcommand{\SIApp}{\emph{SI Appendix}}
\newcommand{\MM}{\hyperref[mm]{\emph{Materials and Methods}}}
\renewcommand{\pi}{\textrm{\greektext p}}
\let\Lambda\varLambda
\let\Sigma\varSigma
\newcommand{\figref}[2]{(Fig.~\hyperref[#1]{\ref*{#1}#2})}
\newcommand{\figrefi}[2]{(Fig.~\hyperref[#1]{\ref*{#1}#2}, inset)}
\newcommand{\textfigref}[2]{Fig.~\hyperref[#1]{\ref*{#1}#2}}
\renewcommand{\eqref}[1]{Eq.~\textbf{\ref{#1}}}
\newcommand{\neqref}[1]{\textbf{\ref{#1}}}
\hypersetup{hidelinks}
\begin{document}
\setlength{\abovedisplayskip}{4pt}
\setlength{\belowdisplayskip}{4pt}
\renewcommand{\floatpagefraction}{.999}
\maketitle
\thispagestyle{firststyle}
\ifthenelse{\boolean{shortarticle}}{\ifthenelse{\boolean{singlecolumn}}{\abscontentformatted}{\abscontent}}{}

\dropcap{T}he folding of tissues into three-dimensional shapes is a crucial part of morphogenetic processes such as neurulation and gastrulation in natural embryonic development~\cite{lowery04,leptin05,keller11,vija17}, associated birth defects such as spina bifida~\cite{copp12}, and synthetic morphogenesis in organoids~\cite{sato09,eiraku11,sasai12,perez-gonzalez21}. These dramatic shape changes are regulated by a complicated interplay of mechanical forces and molecular signalling~\cite{lecuit07,oates09,howard11,lecuit11,heisenberg13,bhide21,moon22,shi22}. The latter can be visualised by fluorescent tagging~\cite{specht17,he19,grimm22}, but the mechanical forces must be inferred indirectly and have therefore remained one of the biggest mysteries in the life sciences.

These mechanical forces are often caused by cell shape changes, including local cell wedging through apical or basal constriction~\cite{lecuit07,keller11,martin20}, which locally imparts a preferred, intrinsic curvature to the tissue. This intrinsic curvature drives tissue bending by generating out-of-plane forces. Such forces not only cause morphogenetic changes, but also elicit biochemical responses~\cite{blonski21}. However, physically, the mechanics and, more biologically, the role in development of different cell shape changes are still unclear: For example, what is the relative contribution to tissue folding of areas of cell wedging and adjacent areas with less pronounced cell shape changes~\cite{bhide21}? Moreover, the local intrinsic curvature driving tissue folding is not in general compatible with the global geometry of the tissue. The folding of the tissue does not in general resolve this geometric incompatibility, so it leads to persisting stresses in the tissue termed residual stresses~\cite{erlich15,ambrosi19}. Now tissues are known to have the ability to alleviate tensile~\cite{iyer19}, compressive, or bending~\cite{fouchard20} stresses through cell movements or changes in cell adhesion or shape and this adaptation can be crucial for maintaining tissue integrity~\cite{acharya18}. How the mechanical state of the tissue is involved in this adaptation in general remains, however, an open question. Addressing these outstanding problems fundamentally relies on quantifying the spatio-temporal distribution of mechanical forces in tissues.   

Over the past decade, different techniques to infer mechanical forces in tissues have therefore emerged, ranging from atomic force microscopy, traction force microscopy, and fluorescence resonance energy transfer-based molecular force sensors to optical tweezers, micropipette aspiration, magnetic beads, and liquid droplets~\cite{gomezgonzalez20,valet22}. However, methods for quantifying out-of-plane stresses in folding epithelia are still lacking. In fact, we are only aware of a single quantitative study of out-of-plane stresses in cell sheets: Fouchard \emph{et al.}~\cite{fouchard20} measured the force needed to unfurl curled synthetic cell sheets and went on to estimate the active torques and the bending modulus of epithelial monolayers~\cite{recho20}. In particular, while numerous studies have used laser ablation to infer in-plane forces from quantifications of the ensuing recoil of the tissue~[see, e.g., Refs.~\cite{shivakumar16,jain20,dye21,marshall22}], extending these methods to out-of-plane forces in geometrically more complex curved dynamic tissues has proved challenging because of the difficulty of imaging out-of-plane recoils on ablation and the need for a mechanical model of these complex deformations to link the measured recoil to the mechanical forces causing it.

Here, we present our framework combining orthogonal laser ablation (OLA, \textfigref{fig1}{A}) and a mechanical model to quantify out-of-plane stresses and infer mechanical properties of curved cell sheets, which we apply to the gastrulation-like inversion process of the microalga \emph{Volvox}.

Inversion in \emph{Volvox} and the related volvocine algae (Chlorophyta) is an emerging model developmental event, during which the spherical embryonic cell sheet turns itself inside out through a programme of cell shape changes~\cite{viamontes77, kirkreview, hallmann06, iida11, iida13, herron16, hohn11, hohn16, matt16}. Here, we focus on type-B inversion in \emph{V. globator}~(\citenum{zimmermann25,hallmann06,hohn11,hohn15}, \mbox{\textfigref{fig1}{B--G,B$'$--G$'$}}). We have previously described this inversion with a mechanical model~\cite{hohn15,haas15,haas18a} in which the programme of cell shape changes driving the process appears as variations of the intrinsic curvatures and intrinsic stretches of an elastic shell~\cite{hohn15,haas15} and which reproduces the inversion process quantitatively~\cite{haas18a}. These \emph{in silico} approaches have made a number of mechanical predictions~\cite{hohn15,haas18a}, but the elastic framework underlying them has in fact remained untested.

We start by using our ablation framework to reveal persistent residual out-of-plane stresses in the posterior hemisphere throughout inversion by showing how the cell sheet unfurls at the boundary of a circular ablation at the posterior pole. We use our mechanical model to show that the geometric incompatibilities causing this recoil result from mismatched intrinsic curvatures of the cell sheet. From quantifications of the recoil, we infer mechanical properties of the posterior hemisphere during invagination and its intrinsic curvature, which is consistent with the observed cell shapes changes. These results do not only therefore provide proof-of-principle of our ablation framework, but also test the elastic framework underlying our previous \emph{in silico} approaches. We go on to extend our elastic model~\cite{haas18a,haas21} to reproduce the tissue shapes quantitatively at different stages of inversion. Strikingly, the resulting fitted mechanical sequence and the ablation data show a clear signature of out-of-plane stresses in the posterior hemisphere and suggest an adaptive response to these stresses. This shows that the tissue dynamics of \emph{Volvox} inversion rely not only on cell wedging, but also on out-of-plane stresses from the additional cell shape changes in the posterior hemisphere.

\subsection*{Type-B inversion in \emph{Volvox globator}} We close this introduction with a more detailed description of type-B inversion in \emph{Volvox globator}, summarising results that this paper relies on. After completion of an initial cell cleaveage phase~\cite{green81a, eheyde22}, the embryos of \emph{V. globator} consist of a spherical monolayer of $\sim3000$ cells~\cite{hallmann06}. Each embryo is located within a fluid-filled embryonic vesicle that is embedded in the extracellular matrix (ECM) of the parent~\cite{kirk86, hallmann03, bheyde20, bheyde23}. The embryonic cells are connected to each other by a network of cytoplasmic bridges (CBs) resulting from incomplete cytokinesis~\cite{hoops06}. A ring of cells at the anterior pole lacks CBs, leaving an opening, the phialopore. During the morphogenetic process of inversion, the embryos turn themselves inside-out through this phialopore within $\sim60-80\,\text{min}$ in order to expose the two cilia that grow at each apical cell pole~\cite{hohn11}.

During inversion the embryonic cells, which have a microtubule-based cortical cytoskeleton, but neither cell walls nor ECM~\cite{kirkbook}, remain connected to each other by CBs, so the positions of the cells relative to their neighbours are maintained~\cite{hohn11}, which suggests our elastic description of the cell sheet~\cite{hohn15,haas15,haas18a,haas18b,haas21}. The global deformations of the embryonic cell sheet during inversion are driven by cell shape changes~\figref{fig1}{B--G} that have previously been described based on electron and confocal microscopy~\cite{hohn11}.

Prior to inversion, all embryonic cells are teardrop-shaped and connected at their broadest point by CBs~\figref{fig1}{B,B$'$}. Type-B inversion begins (``early invagination'') with the appearance of a circular bend region (BR) just below the equator of the embryo, caused by cell-wedging and re-location of the CBs to the thinner basal cell poles~\cite{hohn11}. Simultaneously, the cells in the posterior hemisphere undergo actin-dependent thinning~\cite{nishii99,nishii03} and become spindle-shaped~\figref{fig1}{C,C$'$}. Our previous \emph{in silico} analyses~\cite{hohn15,haas18a} indicate that active contraction of the posterior cell sheet, associated with this latter cell shape change, is necessary to explain the resulting ``mushroom shape'' of the embryos. As inversion proceeds (``late invagination''), a wave of cell wedging travels from the equator towards the posterior pole and the posterior hemisphere moves into the anterior one~\cite{hohn11}. The cells at the posterior pole retain their spindle shape~\figref{fig1}{D,D$'$}. The posterior hemisphere then moves entirely into the anterior, but a ``dimple'' remains where the cell sheet around the posterior pole has not yet inverted. The cells at the posterior pole remain spindle-shaped while the basal cell poles in the BR are less wedge-shaped than before, resulting in a relaxation of the curvature in the BR and near the posterior pole~\figref{fig1}{E,E$'$}. By the time the posterior hemisphere has fully inverted, the phialopore has widened and the anterior hemisphere has started to peel over the inverted posterior. The cells in the inverted posterior have adopted a pencil shape~\figref{fig1}{F,F$'$} at this stage. Closure of the phialopore marks the completion of inversion and the end of embryogenesis in \emph{Volvox}. At this stage, all cells are pencil-shaped with pointy apical cell poles~\cite{hohn11}. Within $30-60\,\text{min}$ after inversion, all cells shorten, widen, and adopt a columnar shape while the radius of the juvenile spheroid increases~\figref{fig1}{G,G$'$}.

\section*{Experiments}
\subsection*{Orthogonal laser ablation (OLA) in type-B \emph{Volvox} inversion}
To quantify out-of-plane stresses in \emph{Volvox} inversion, we performed laser ablation experiments on \emph{V. globator} embryos in consecutive stages of inversion (\MM). In order to capture out-of-plane elastic responses to these ablations, they were performed on the mid-sagittal cross-sections of the axisymmetric \emph{V. globator} embryos. Cross-sections were imaged using 2-photon microscopy (\textfigref{fig1}{A} and \MM), and approximately circular holes were created at the posterior pole of the embryos by laser ablation using a separate 2-photon laser (\MM). The axisymmetry of the embryos and the ablations ensures that any subsequent deformation of the cell sheet is approximately axisymmetric, too. The dimensions of the resulting hole in the cell sheet were determined in three-dimensional datasets recorded subsequently (\MM).

\begin{figure*}[h!]
\vspace{-3mm}
\centering\includegraphics{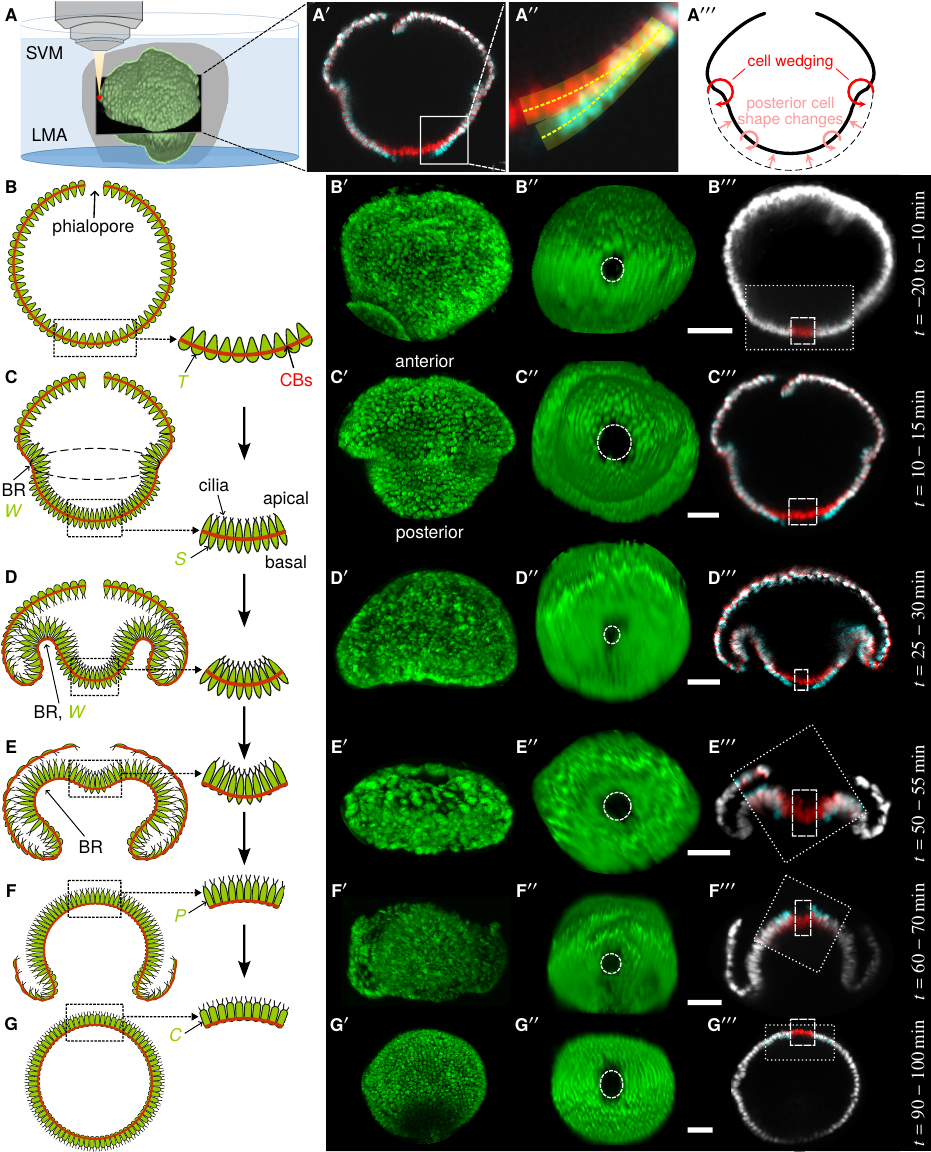}
\caption{Experiments: orthogonal laser ablation (OLA) of inverting \emph{Volvox globator} embryos. (A) Experimental setup for 2-photon microscopy and OLA, illustrated for a surface rendering of a \emph{V. globator} embryo at an early inversion stage. Samples were embedded in low-melting point agarose (LMA), submerged in Standard \emph{Volvox} Medium (SVM), and imaged using 2-photon microscopy. Holes were created using a separate 2-photon laser (\MM) at the posterior pole of embryos before, during, and after inversion. (A$'$) Overlay (white) of micrographs of the mid-sagittal cross-section of the embryo in panel (A), just before ablation (red) and after relaxing into its post-ablation equilibrium shape (cyan), showing the unfurled edge of the cell sheet. (A$''$) Tracing of the cell sheet (transparent area) and its midline (dashed line) in the square magnified from panel (A$'$). (A$'''$) Mechanical hypothesis: posterior cell shape changes not only contract the posterior hemisphere~\cite{hohn15}, but also induce a curvature mismatch in it that causes out-of-plane stresses distant and distinct from those generated by the wedge-shaped cells in the bend region. (B)--(G) Schematic cross-sections of \emph{V. globator} embryos in consecutive developmental stages, modified from Ref.~\cite{hohn11}, showing associated cell shapes, the network of cytoplasmic bridges (CBs, red line), and the growing cilia. BR: bend region, W: wedge-shaped cells. Insets show cell shapes~\cite{hohn11} near the posterior pole of the embryo (T: teardrop-shaped cells; S: spindle-shaped cells; P: pencil-shaped cells; C: columnar cells). (B$'$)--(G$'$) Maximum intensity projections of z-stacks showing lateral views of \emph{V. globator} embryos in the developmental stages in panels (B)--(G). The anterior and posterior hemispheres are labelled in panel (C$'$). (B$''$)--(G$''$) Posterior view of the embryos in panels (B$'$)--(G$'$) after laser ablation. Dashed white line: outline of the hole created by laser ablation after relaxation into its post-ablation equilibrium shape. (B$'''$)--(G$'''$) Overlays (white) of micrographs of the mid-sagittal cross-section of the embryos in panels (B$'$)--(G$'$) just before ablation (red) and after relaxing into their post-ablation equilibrium shapes (cyan). Dashed lines: areas of laser ablation. Thin dotted lines in panels (B$'''$), (E$'''$)--(G$'''$): field of view in which the laser ablation was performed. Times $t$ are given relative to the start time $t=0$ of inversion. Scale bars: $20\,\text{\textmu m}$.\vspace{-24pt}
}\label{fig1} 
\end{figure*}

\subsection*{Observations}
Laser ablations were performed at the posterior pole of embryos in consecutive developmental stages before, during, and after inversion~\figref{fig1}{B$''$--G$''$,B$'''$--G$'''$}. 

Prior to inversion, embryos did not show any recoil on laser ablation~\figref{fig1}{B$'$--B$'''$}, i.e. no deformation within at least $2\,\text{min}$ following ablation ($N = 5$ embryos), indicating that the cell sheet is not residually stressed before the onset of inversion. However, during early invagination, ablation led to outwards unfurling of the edges of the cell sheet and relaxation into a new equilibrium shape~(\textfigref{fig1}{C$''$,C$'''$}; $N = 14$), showing that out-of-plane residual stresses have appeared in the cell sheet. Ablations during the late invagination stage again led to unfurling of the edges of the cell sheet~(\textfigref{fig1}{D$''$,D$'''$}; $N = 5$). When ablations were performed at the ``dimple stage''~\figref{fig1}{E,E$'$}, the unfurling cell sheet
adopted a curvature closer to that in the BR~(\textfigref{fig1}{E$''$,E$'''$}; $N = 5$), indicating a mechanical effect of the BR on the post-ablation equilibrium shape. Interestingly, for ablations at the stage where the posterior hemisphere has fully inverted~\figref{fig1}{F,F$'$}, the direction of unfurling changed compared to earlier timepoints~(\textfigref{fig1}{F$''$,F$'''$}; $N = 7$). Finally, for posterior ablations after the cells have become columnar~\figref{fig1}{G,G$'$}, the cell sheet no longer shows any recoil within at least $2\,\text{min}$ after laser ablation~(\textfigref{fig1}{G$''$,G$'''$}; $N = 4$), indicating that the residual stresses have relaxed.

At all stages of inversion observed, unfurling of the edges of the cut upon laser ablation and relaxation into the post-ablation equilibrium shape took approximately $4-10\,\text{s}$. Within this time, the embryos otherwise maintained their shape: Morphogenetic inversion movements, i.e. global embryonic shape changes distinct from the local ablation response, were only distinguishable after at least $1\,\text{min}$. While the hole caused by laser ablation did not close up, the embryos did complete inversion after ablation, unless the cut size exceeded about half the radius of the posterior hemisphere in which case the anterior hemisphere failed to invert (not shown).  

The observed outward unfurling of the cell sheet suggests that the preferred, intrinsic curvature of the cell sheet differs from the curvature into which its unablated spherical shape forces the cell sheet. We therefore hypothesise that such a curvature mismatch gives rise to out-of-plane stresses in the posterior hemisphere, including stresses distant and hence distinct from the out-of-plane stresses in the bend region of wedge-shaped cells~\figref{fig1}{A$'''$}. The thinning of cells from teardrop to spindle shapes driving contraction of the posterior hemisphere also changes its curvature~(\citenum{haas15}, \textfigref{fig1}{A$'''$,B,C}), so would contribute to this mismatch of curvatures. The change of the direction of unfurling with the change of the sign of the curvature of the posterior hemisphere~\figref{fig1}{B$'''$,F$'''$} is consistent with this hypothesis if the intrinsic curvature of the tissue is smaller in magnitude than its actual curvature.

\section*{Origin of out-of-plane stresses in \emph{Volvox} inversion}
\subsection*{Mechanical toy problems} To test this hypothesis, we need to understand the mechanics of ablations in curved tissues. For this purpose, we introduce three mechanical toy problems.
\subsubsection*{Toy problem 1: deformations of a spherical elastic shell with mismatched intrinsic curvatures} 
\begin{SCfigure*}[\sidecaptionrelwidth][h!]
\includegraphics{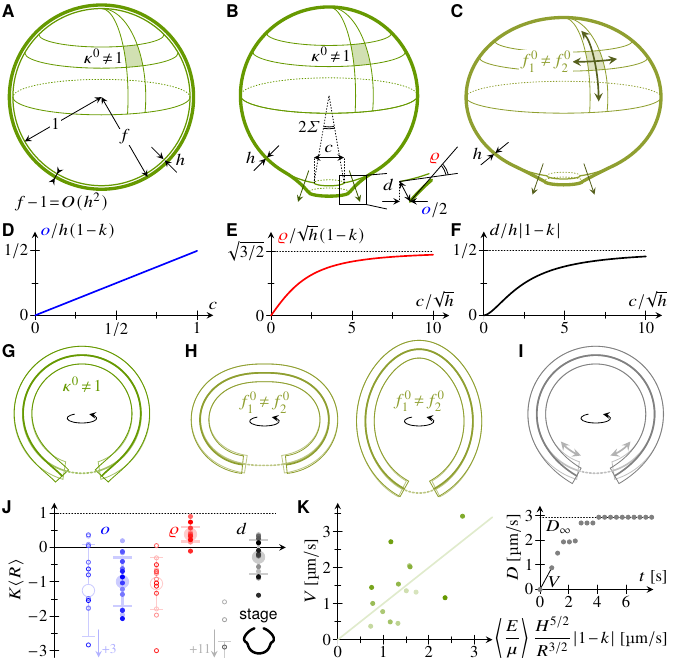}
\caption{Origin of out-of-plane stresses in \emph{Volvox inversion}: mechanical toy problems and analysis of experiments. (A)~Toy problem~1: a spherical shell of unit radius and  thickness $h\ll1$ deforms to a radius $f=1+O(h^2)$ due to mismatched intrinsic curvatures $\kappa^0=k\neq1$. (B)~Toy problem~2: this mismatch of intrinsic curvatures causes a circular cut of angle $2\Sigma$ and radius $c=2\sin{\Sigma}$ to open. Inset: definition of the opening~$o$, rotation $\varrho$, and displacement $d$ of the cut edge. (C)~Toy problem~3: anisotropic contraction also causes recoil on ablation. (D)~Plot of $o\sim h(1-k)$ against~$c$. (E) Plot of asymptotic approximation for $\varrho/\sqrt{h}(1-k)$ against $c/\sqrt{h}$. (F) Plot of asymptotic approximation for $d/h(1-k)$ against $c/\sqrt{h}$. (G) Numerical solution of toy problem~2: plot of cross-sections pre-ablation (light) and post-ablation (dark), showing a recoil qualitatively consistent with experiments. (H) Numerical solutions of toy problem~3: the recoil is much smaller than in experiments despite the excessive anisotropy. (I)~Fast localised relaxation can also reproduce the experimental recoil qualitatively. (J)~Estimated dimensional intrinsic curvatures $K$, nondimensionalised with the average posterior radius $\langle R\rangle$ during invagination. Estimates computed from asymptotic results (open markers) and numerical solutions of toy problem~2 (filled markers) using measurements of $o$, $\varrho$, $d$ for $N=14$ \emph{Volvox} embryos at early invagination (inset). Arrows: off-scale values; large markers and error bars: mean and standard deviation of estimates. (K)~Plot of the corresponding initial recoil velocity $V$ against its scaling. Variables: elastic modulus $E$, effective viscosity~$\mu$, (dimensional) cell sheet thickness $H$, posterior radius $R$ during invagination, (dimensionless) estimated intrinsic curvature~$k$. A straight-line fit estimates the mean $\langle E/\mu\rangle$. Inset: example measurement of the (dimensional) cut displacement~$D$ against time, showing the initial velocity and equilibrium displacement $D_\infty$ used for estimates.} \label{fig2}
\end{SCfigure*}

Our first toy problem studies the effect of curvature mismatch on cell sheet shapes before ablation: we consider a complete spherical shell of unit radius and (relative) thickness $h\ll 1$. The intrinsic curvatures of the shell are $\kappa^0=k\not=1$, different from the undeformed curvature of the shell. This curvature mismatch causes the shell to deform; we assume that the shell remains spherical, and denote by $f$ its deformed radius~\figref{fig2}{A}. 

This radius is determined by the competition of the strains $E=f-1$ due to the stretching of the shell and the bending strains $K=1/f-k$ resulting from the difference of the actual and intrinsic curvatures of the shell. The elastic energy of the shell is the sum of its stretching and bending energies~\cite{landaulifshitz}, so is proportional to
\begin{align}
E^2+\dfrac{h^2}{12}K^2=(f-1)^2+\dfrac{h^2}{12}(1/f-k)^2. \label{eq:shellenergy}
\end{align}
Assuming $k=O(1)$, this is minimised for
\begin{align}
f\sim1-\dfrac{h^2}{12}(k-1). \label{eq:toy1res}
\end{align}
As expected, the shell grows ($f>1$) or shrinks ($f<1$) if its preferred curvature is flatter ($k<1$) or larger (${k>1}$) than its undeformed curvature, but the deformations are asymptotically small compared to the shell thickness.

There is an interesting mechanical subtlety of this calculation: Though this argument gives the correct $O(h^2)$ scaling for the deformations of the shell, the smallness of the deformations is beyond the realm of validity of the shell theory assumed in writing down~\eqref{eq:shellenergy}. For this reason, the correct prefactor has to be calculated within bulk nonlinear solid mechanics; the calculation is given in \SIApp, where we also discuss some of the suprising behaviour revealed by that prefactor which shows how even seemingly innocuous problems in nonlinear mechanics, such as this one, can break our intuition.

\subsubsection*{Toy problem 2: circular ablation of a spherical elastic shell with mismatched intrinsic curvatures}
Next, we study the mechanics of a circular ablation in a shell with mismatched intrinsic curvatures. The mismatched intrinsic curvatures and the no-torque boundary condition at the rim of the cut cause the shell to deform near the cut. For an ablation of angle $2\Sigma$, i.e. radius $c=2\sin{\Sigma}$~\figref{fig2}{B}, asymptotic solution of the equations of shell theory for $h\ll 1$ (\SIApp) shows that the cut opens by an amount~\figref{fig2}{D}
\begin{align}
o\sim h(1-k)\sin{\Sigma}=\dfrac{h}{2}c(1-k).
\end{align}
Importantly, this is asymptotically larger than the deformation without ablation described by \eqref{eq:toy1res}. Asymptotic approximations can also be found (\SIApp) for the rotation $\varrho$ of the rim of the cut and its displacement $d$ due to the ablation~\figref{fig2}{E,F}.

\subsubsection*{Toy problem 3: circular ablation of a spherical elastic shell with anisotropic contraction} Opening of an ablation does not require intrinsic curvature mismatches, however: As explained in more detail in \SIApp, geometry requires the stretches of an axisymmetric spherical shell to be equal at its poles, so anisotropic contraction will cause stresses there and hence result in a recoil on ablation~\figref{fig2}{C}.

\subsection*{Discussion} These results allow us to test our hypothesis, that there is a curvature mismatch during \emph{Volvox} inversion. First, toy problems 1 and 2 emphasise that ablation experiments are needed: The unablated deformations are too small to be visualised in experiments; only the deformations from ablations are comparable to the thickness of the cell sheet, so allow quantification of curvature mismatches that cannot be quantified from the unablated cell sheet only. 

\subsubsection*{The observed recoil during \textit{Volvox} invagination results from a curvature mismatch} Next, solving for post-ablation shapes in toy problems 2 and 3 numerically (\MM), we observe that the recoil for an intrinsically flat shell ($k=0$) agrees with experiments qualitatively~(\textfigref{fig1}{B$'''$}, \textfigref{fig2}{G}). By contrast, the recoil resulting from anisotropic contraction is much smaller than the experimentally observed one, even for blatantly excessive anisotropies~(\textfigref{fig1}{B$'''$}, \textfigref{fig2}{H}). This does not yet allow the conclusion that the observed recoil results from a curvature mismatch: Indeed, fast biological processes could \emph{a priori} be triggered close to the ablation site to lead to a fast active relaxation of the shell there. Indeed, this could result in a geometric incompatibility leading to a recoil qualitatively consistent with experiments (\textfigref{fig2}{I} and \MM). If however this mechanism underlied the experimental observations, we would expect to observe a recoil for pre- and post-inversion ablations, too, which we do not~\figref{fig1}{B$'''$,G$'''$}. We therefore discard this possibility, and conclude that the observed recoil is only consistent with a curvature mismatch that existed before ablation.
\subsubsection*{Quantitative estimates of the curvature mismatch during \textit{Volvox} invagination are consistent with the observed cell shape changes} To make this analysis more quantitative, we measured the recoil on ablation in $N=14$ \emph{Volvox} embryos during early invagination (\textfigref{fig1}{C--C$'''$} and \MM), and extracted estimates of intrinsic curvatures \figref{fig2}{J} from the measurements of the cut opening $o$, rotation $\varrho$, and displacement $d$ using the asymptotic approximations \figref{fig2}{D--F} and numerical solutions of toy problem 2. The hemispherical shape of the posterior at these early inversion stages \figrefi{fig2}{J} justifies using toy problem 2 for this inference. The estimates based on asymptotic approximations have a much larger variance than those based on numerical solutions~\figref{fig2}{J}. This suggests that, while the asymptotic estimates are useful to understand the mechanical basis of the recoil upon ablation, the cell sheet of \emph{Volvox} is too thick for them to yield good quantitative estimates. Meanwhile, we observe that the estimate, based on numerics, of mean intrinsic curvature from $o$ has a larger standard deviation than the estimates from $\varrho$ and $d$~\figref{fig2}{J}. Moreover, the latter are within one standard deviation of each other, while the former is not~\figref{fig2}{J}. This is perhaps not unexpected, as measuring $o$ requires quantifying smaller deformations. We therefore estimate the mean intrinsic curvature of the cell sheet based on $\varrho$ and $d$ only, finding $\langle k\rangle \approx 0.05\ll 1$. Importantly, this estimate is consistent with the observed cell shape changes: The spindle-shaped cells in the posterior during early \emph{Volvox} invagination are symmetric, and connected at their midplane by cytoplasmic bridges~\figrefi{fig1}{C}, suggesting $k=0$. By contrast, the teardrop-shaped cells in preinversion embryos~\figrefi{fig1}{B} suggest a positive intrinsic curvature, and indeed the absence of recoil observed preinversion (i.e. the absence of residual stresses) requires $k=1$.
\subsubsection*{Recoil velocity measurements yield a bound on the elastic modulus of the cell sheet} The (dimensional) recoil velocity $V$ is set by the interplay between the elastic force driving it and viscous dissipation in the tissue and surrounding fluid. A scaling argument (\SIApp) yields
\begin{align}
V\sim \dfrac{E}{\mu}\dfrac{H^{5/2}}{R^{3/2}}|1-k|, 
\end{align}
in which $E$ is the elastic modulus of the cell sheet, $\mu$ is an effective viscosity, and $H$ and $R$ are the dimensional thickness and posterior radius of the cell sheet. We obtain $V$ from measurements of the initial recoil velocity of the cell sheet~\figref{fig2}{K}. We stress our use of this initial velocity for these dynamic measurements, by contrast with the above matching of (quasi-)equilibrium quantifications of the recoil to a static toy problem~\figrefi{fig2}{K}. From the scaling relation, we obtain a mean value $\langle E/\mu\rangle\approx 0.6\,\mathrm{s}^{-1}$~\figref{fig2}{K}. With the lower bound $\mu\gtrsim 0.8\,\mathrm{mPa\,s}$ corresponding to the viscosity of algal growth medium~\cite{petkov96}, so neglecting dissipation within the tissue, this yields $E\gtrsim 0.5\,\mathrm{mPa}$, much lower than the values reported for confluent tissues, yet perhaps appropriate for an extremely floppy, non-confluent tissue. With $H\approx 11\,\text{\textmu m}$~\cite{hohn11}, the bending modulus $K=EH^3\gtrsim 7\,\cdot10^{-19}\,\mathrm{J}$, well above the energy of thermal fluctuations, provides a sanity check of this estimate.
\begin{figure*}[h!]
\includegraphics{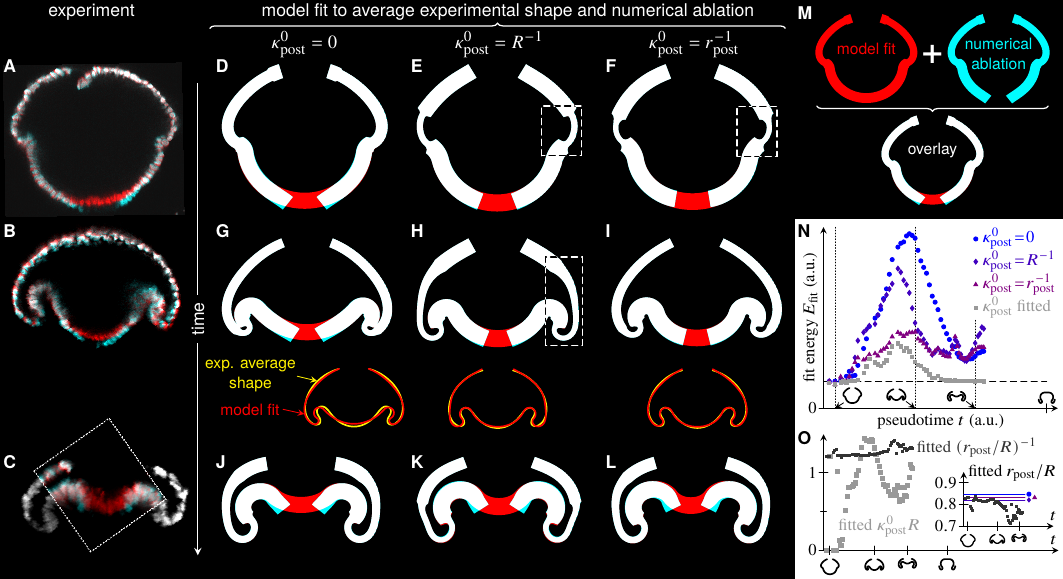}
\caption{Effect of out-of-plane stresses in \emph{Volvox} inversion: fits of the elastic model to average experimental cross-sections~\cite{haas18a} and numerical ablations. (A)~Overlay of experimental cross-sections (white) before (red) and after (cyan) a posterior ablation at an early invagination stage~\figref{fig1}{B$'''$}. (B)~Analogous plot for an ablation at a late invagination stage~\figref{fig1}{C$'''$}. (C)~Analogous plot for an ablation at a mid-inversion stage~\figref{fig1}{D$'''$}. Outside the dotted rectangle, only pre-ablation shapes are plotted (white). (D)~Fit, for zero posterior intrinsic curvatures $\smash{\kappa^0_{\mathrm{post}}=0}$, of the elastic model to the average experimental cross-section at an inversion timepoint corresponding to panel (A) (red), overlaid with equilibrium shape after numerical ablation (cyan), showing a recoil on ablation. (E)~Analogous plot, for posterior intrinsic curvatures $\smash{\kappa^0_{\mathrm{post}}=R^{-1}}$, equal to the undeformed curvature of the embryo. The recoil is reduced. The dashed box highlights thinning of the cell sheet in the fitted shape. (F)~Analogous plot, for posterior intrinsic curvatures $\smash{\kappa^0_{\mathrm{post}}=r_{\mathrm{post}}^{-1}}$, equal to the inverse contracted intrinsic radius of the posterior. There is almost no recoil. (G)~Fit, for $\smash{\kappa^0_{\mathrm{post}}=0}$, of the elastic model to the average experimental cross-section at an inversion timepoint corresponding to panel (B), and numerical ablation. Inset: comparison of fitted shape and average experimental cross-section. (H)~Analogous plot for $\smash{\kappa^0_{\mathrm{post}}=R^{-1}}$. The fitted shape reproduces the experimental shape better than that in panel (G). (I)~Analogous plot for $\smash{\kappa^0_{\mathrm{post}}=r_{\mathrm{post}}^{-1}}$. The fit is again better than that in panel (G). (J)~Fit, for $\smash{\kappa^0_{\mathrm{post}}=0}$, of the elastic model to the average experimental cross-section at an inversion timepoint corresponding to panel (C), and numerical ablation. (K)~Analogous plot with $\smash{\kappa^0_{\mathrm{post}}=R^{-1}}$. The recoil on ablation is qualitatively similar. (L)~Analogous plot with $\smash{\kappa^0_{\mathrm{post}}=r_{\mathrm{post}}^{-1}}$. (M)~Illustration of the numerical ablation plots using the plots in panel (D): the superposition of the fitted shape before numerical ablation (red) and that after numerical ablation (cyan) defines the overlaid shape (white). (N)~Plot of the fit energy $E_{\mathrm{fit}}$ (arbitrary units) against inversion (pseudo)time for fits with $\smash{\kappa^0_{\mathrm{post}}\in\{0,R^{-1},r_{\mathrm{post}}^{-1}}\}$ and for a fit in which $\smash{\kappa^0_{\mathrm{post}}},\smash{r_{\mathrm{post}}}$ are among additional fitting parameters. $E_{\mathrm{fit}}=0$ would indicate a perfect fit; the horizontal dashed line indicates the value of $E$ at which fitting was cut off. Pseudotimepoints used for the fits in panels (D)--(L) are highlighted. (O)~Plots of fitted $\smash{\kappa^0_{\mathrm{post}}}R$ and $\smash{(r_{\mathrm{post}}/R)^{-1}}$ against inversion (pseudo)time. Inset: plot of fitted $\smash{r_{\mathrm{post}}}/R$ against inversion pseudotime. The constant values used for the fits with $\smash{\kappa^0_{\mathrm{post}}\in\{0,R^{-1},r_{\mathrm{post}}^{-1}}\}$ are highlighted.}\vspace{-4pt}\label{fig3}
\end{figure*}
\section*{Effect of out-of-plane stresses in \emph{Volvox} inversion}
Having thus shown that out-of-plane stresses resulting from a curvature mismatch persist during \emph{Volvox} invagination, we ask: What is the effect of these out-of-plane stresses on the tissue shape sequence during inversion?
\subsection*{Quantitative model of \emph{Volvox} inversion}
The toy models in the previous section can describe the shapes of \emph{Volvox} embryos during early invagination locally, close to the posterior ablation, but cannot capture the global embryo shapes and mechanics. We therefore extended our previous detailed morphoelastic model of the cell shape changes of inversion~\cite{haas18a} and fitted parameters encoding the cell shape changes driving inversion~\cite{hohn11} to the average cell sheet midlines at various timepoints of inversion~\cite{haas18a}, as described in detail in \MM{} and \SIApp. We then performed numerical ablations on these fitted shapes and compared \emph{in vivo} and \emph{in silico} results~\figref{fig3}{A--M}.
\subsubsection*{Results for early invagination stages are consistent with the observed cell shape changes and the estimates from ablation recoils} We first considered an early invagination timepoint~\figref{fig3}{A}, for which we fitted the average embryo shape for three different values of the posterior intrinsic curvature $\kappa_{\mathrm{post}}^0$~\figref{fig3}{D--F}: $\kappa_{\mathrm{post}}^0=0$, corresponding to the flat posterior intrinsic curvature predicted by the quantitative estimates of the previous section and the observed cell shape changes in the posterior; $\kappa_{\mathrm{post}}^0=R^{-1}$, equal to the inverse radius of the undeformed embryo, i.e. the value suggested at pre-inversion stages by the lack of recoil on ablation; $\kappa_{\mathrm{post}}^0=r_{\mathrm{post}}^{-1}$, equal to the inverse intrinsic contracted radius of the posterior resulting from the cell shape changes to spindle-shaped cells~\figref{fig1}{C}. With this last value, the contracted shape of the uninverted posterior is unstressed. The model can reproduce the average embryo midlines well for all three values of $\kappa_{\mathrm{post}}^0$~\figref{fig3}{D--F}, although the fit parameters inferred from the cell sheet midlines for the latter two values imply a thinning of the anterior hemisphere of the cell sheet~(highlighted in \textfigref{fig3}{E,F}) that is only observed experimentally at slightly later stages of invagination, when it associated with the formation of pancake-shape cells there~(\citenum{hohn11}, \textfigref{fig1}{D}). However, numerical ablations reveal that these three fitted shapes are mechanically different~\figref{fig3}{D--F}: Only the first case leads to a sizeable recoil on ablation, qualitatively consistent with experiments. The fitting thus suggests that $\kappa_{\mathrm{post}}^0=0$ and is therefore consistent with the estimates of the previous section and the observed cell shape changes.
\subsubsection*{Results for late inversion stages are consistent with the observed expansion of the region of cell wedging} Next, we considered a late inversion timepoint~\figref{fig3}{C}, around the time of phialopore opening. Again, all three values of posterior intrinsic curvature could reproduce the observed cell sheet midlines~\figref{fig3}{J--L} and the fitted shapes even reproduce the observed thinning of the anterior hemisphere~\cite{hohn11} even though the fits are based on the cell sheet midlines only. Moreover, numerical ablations lead to similar recoils in all three cases~\figref{fig3}{J--L}. This is consistent with the observed expansion of the wave of cell wedging towards the posterior pole~\cite{hohn11}: The recoil magnitude is set no longer by the geometric incompatibilities of the few remaining spindle- or pencil-shaped cells~\figrefi{fig1}{E}, but by that of the wedge-shaped cells in the bend region~\figref{fig1}{C}.
\subsection*{Discussion}
These results thus provide a check of the fitting and suggest that the mechanics of out-of-plane stresses in \emph{Volvox} inversion can be captured correctly by this fitting. In turn, this indicates that we can use the fit results to understand out-of-plane stresses at other stages of inversion.
\subsubsection*{The fitted mechanics at midinversion stages predict changing posterior out-of-plane stresses} We thus turn to a midinversion stage next~\figref{fig3}{B}, again fitting the model to the experimental data for the three different values of $\kappa_{\mathrm{post}}^0$~\figref{fig3}{G--I}. Interestingly, the quality of the fit to the average experimental inversion shape is considerably worse for $\kappa_{\mathrm{post}}^0=0$~\figref{fig3}{G, inset} than for the other two values~\figref{fig3}{H--I, insets}. Interestingly, the fit parameters inferred from the cell sheet midlines in the case $\kappa_{\mathrm{post}}^0=R^{-1}$ imply a thickness gradient in the anterior hemisphere~\figref{fig3}{H} that is consistent with the formation of pancake-shaped cells starting in the anterior fold and propagating towards the phialopore, as described previously~(\citenum{hohn11}, \textfigref{fig1}{D}). All of this suggests that the intrinsic curvature of the spindle-shaped cells and hence the out-of-plane stresses in the uninverted posterior change at midinversion stages. The fact that a better fit~\figref{fig3}{H--I} is obtained for positive values of $\kappa_{\mathrm{post}}^0$ that are associated with reduced geometric incompatibilities in the uninverted posterior begs the question: Could this change constitute an adaptive response of the tissue to the stresses resulting from the geometric incompatibility generated by the spindle-shaped cells in the posterior?
\subsubsection*{The fitted mechanical sequence hints at an adaptive response to out-of-plane stresses} To provide a hint of an answer to this question, we considered fits of our model to the experimental data for more developmental timepoints before phialopore opening\footnote{We did not fit our model beyond the onset of phialopore opening because that is associated with poorly understood cell rearrangements near the phialopore~\cite{haas18a}, the mechanics of which cannot be described by our elastic model.} and for the three values of $\kappa_{\mathrm{post}}^0$. While the fit energies~\figref{fig3}{N} are comparable at early and late stages of inversion, those at mid-inversion stages are lowest, by some margin, for the fit with $\kappa_{\mathrm{post}}^0=r_{\mathrm{post}}^{-1}$. We have already noted above that the contracted shape of the uninverted posterior is unstressed in this case. This therefore hints that $\kappa_{\mathrm{post}}^0$ changes in such a way as to relieve the geometric incompatibilities in the uninverted posterior, and hence at an adaptive response to the stresses. To substantiate this hint, we performed a further fit in which $\kappa_{\mathrm{post}}^0,r_{\mathrm{post}}$ are among additional fitting parameters~(\SIApp). The resulting fit energy~\figref{fig3}{N} is not substantially lower than that of the more constrained fit with $\kappa_{\mathrm{post}}^0=r_{\mathrm{post}}^{-1}$ at mid-inversion stages. Moreover, the fitted values of $\smash{\kappa_{\mathrm{post}}^0}$~\figref{fig3}{O} increase from zero at the start of invagination to values comparable to $r_{\mathrm{post}}^{-1}$ at mid-inversion stages. All of this lends further support to the hypothesis of an adaptive response to stresses in the tissue. Interestingly, the fitted values of $\kappa_{\mathrm{post}}^0$~\figref{fig3}{O} decrease and those of $r_{\mathrm{post}}$~\figref{fig3}{O, inset} decrease before increasing again at mid-to-late inversion stages, when the fit energy of this fit becomes rather smaller than that of the other fits~\figref{fig3}{N}. All of this could be a signature of the pencil-shaped cells~(\citenum{hohn11}, \textfigref{fig1}{E}, inset) that begin to form at these stages of inversion (and another adaptive response): Indeed, the position of cytoplasmic bridges in these cells may suggest $\kappa_{\mathrm{post}}^0R<1$, and they are narrower than spindle-shaped cells~\cite{hohn11}, indicating reduced $r_{\mathrm{post}}$. We discuss these results further in \SIApp.
\section*{Conclusion}
We have combined ablation experiments and a detailed elastic model to establish the origin and analyse the mechanical effects of out-of-plane stresses during \emph{Volvox} inversion. Strikingly, our results show how changes of posterior intrinsic curvature---albeit small compared to the intrinsic curvature imposed by the wedge-shaped cells in the bend region---contribute to the observed tissue shape sequence of \emph{Volvox} inversion. In particular, the theory predicts an intriguing adaptive response of the geometric incompatibilities in the cell sheet to its out-of-plane stresses.

Tissue bending is not therefore all about apical constriction: Rather, out-of-plane stresses from other cell shapes changes also contribute mechanically to \emph{Volvox} inversion. This is possible for biological tissues are not infinitely thin: If they were, the asymptotic separation of stretching and bending energies~\cite{landaulifshitz} would imply that only the large bending deformations of apical constriction~\cite{recho20,haas21} that break this asymptotic separation would have an effect comparable to that of in-plane stretching deformations. The question now becomes: Do these out-of-plane stresses that impact the shapes of inverting \emph{Volvox} embryos and their adaptive response predicted by theory have a mechanical function? Since inversion is still
possible for all three scenarios in \textfigref{fig3}{D--L}, they are not required mechanically for inversion, unlike the formation of wedge-cells or contraction, which is known to be required mechanically for the closely related type-A inversion in \emph{V. carteri}~\cite{nishii99,nishii03,haas18b}. However, one might speculate that these stresses or equivalently the stored elastic energy contribute to the mechanical robustness of \emph{Volvox} inversion. Also, these stresses and their predicted adapation during inversion could have a function related to mechanosensing and hence to the unknown signalling processes that orchestrate inversion. Testing such hypotheses will require again close integration of experiment and theory and, in particular, dynamic quantification of cell shapes to link ``microscopic'' cell-level changes to geometric incompatibilites and hence stresses at the ``macroscopic'' tissue level.

Our experimental and theoretical approach sets out a framework for inferring the mechanical state and properties of curved tissues based on their unfurling on ablation, which we expect to be applicable to a broad range of problems in cell sheet folding. However, it has also revealed key differences between the analysis of ablations in curved and flat tissues: First, the standard measure of recoil on ablation in flat tissues, i.e. what we termed the cut opening above, may not be appropriate for analysing ablations in curved tissues, because it requires quantification of smaller deformations than the measures of cut rotation and cut edge displacement that we introduced here, and is therefore more prone to experimental noise. Second, the inference of cell sheet properties from the deformations on ablations is more complex in curved tissues: It may require numerical solution of a ``toy problem'' rather than use of a closed-form expression because tissues are not asymptotically thin, as already noted above: The tissue thickness, which is the natural small parameter for asymptotic solution of such toy problems, may not be sufficiently small for closed-form asymptotic expressions to enable quantitative inference.

Finally, from a more physical point of view, it is striking that the ``simple'' tissue shapes of early \emph{Volvox} invagination did not constrain the mechanical state of the cell shape to the same extent as the more complex shapes at mid-inversion stages: At early invagination stages, knowledge of the unfurling of the tissue on ablation was required to distinguish between the three mechanical scenarios discussed in \textfigref{fig3}{D--F}, while, at later stages, we could distinguish between these possibilities based on the shapes of cell sheet alone~\figref{fig3}{G--I}. It is tempting to ask whether this observation generalises: Do more complex tissue shapes intrinsically contain more information about the mechanical state of the tissue? Of course, this begs a more basic question: What is the right way of quantifying tissue shape complexity? Addressing these question both in simple physical model problems and in simple biological systems like \emph{Volvox} or synthetic, organoid models will allow us to take the first steps towards addressing, too, the much more fundamental biological problem that is the relation between tissue shape (complexity) and biological function. 

\matmethods{\label{mm}\vspace{-5mm}
\subsection*{Model organism cultivation} Wild-type strain \emph{Volvox globator} Linné (SAG 199.80) was obtained from the Culture Collection of Algae at the University of Göttingen, Germany~\cite{schlosser94}, and cultured as described previously~\cite{brumley14,hohn15} in liquid Standard \emph{Volvox} Medium (SVM) with a cycle of $16\,\text{h}$ light at $24^\circ\text{C}$ and $8\,\text{h}$ dark at $22^\circ\text{C}$. \emph{V. globator} cultures in the asexual life cycle of these monoecious microalgae were used.
\subsection*{2-photon microscopy and laser ablation experiments} \emph{Volvox} spheroids containing embryos undergoing inversion were embedded in $2\,\text{\textmu l}$ of $1\%$ low-melting-point agarose (LMA), covered with SVM, and imaged using a Trim Scope 2-photon microscope (LaVision). A laser line at $\lambda = 1040\,\text{nm}$ was used for imaging and a separate laser line at $\lambda=900\,\text{nm}$ for performing laser ablations. As a single chloroplast mostly fills each \emph{Volvox} cell, chlorophyll-autofluorescence was detected at $\lambda>647\,\text{nm}$, and used for visualising the embryonic cell sheet. Z-stacks were recorded before and after each laser ablation experiment, with a z-step of $2\,\text{\textmu m}$. To maximise acquisition speed and capture the elastic response to the laser cut, a single plane was imaged during ablation experiments, which removed cells within a radius of $3-6\,\text{\textmu m}$ of the posterior pole of the cell sheet. Videos of the mid-sagittal plane were recorded at $1-4.5\,\text{fps}$ for at least $10\,\text{s}$ before and $30\,\text{s}$ after laser ablation. Videos used for determining velocities of elastic recoils were recorded at the maximal speed possible for the field of view size used, corresponding to $2.5-4.5\,\text{fps}$.
\subsection*{Analysis of experimental data} Outlines of the posterior hemisphere were traced manually on recordings of midsagittal embryo cross-sections using \textsc{Fiji}~\cite{fiji}. To trace the midline of the cell sheet, the line width was set to fit the thickness of the cell sheet. Recoil quantification was then performed using custom \textsc{Matlab} (The MathWorks, Inc.) scripts; in particular, the fit in~\textfigref{fig2}{K} uses the \texttt{polyfitZero} function by M. Mikofski, obtained from the \textsc{Matlab} file exchange (file 35401). Posterior cell sheet radii were estimated by fitting a circle to the cross-sections in \textsc{Fiji}. Cell sheet thicknesses were estimated by scaling the chloroplast thicknesses measured in \textsc{Fiji} so that the resulting mean value matched that reported previously~\cite{hohn11}.
\subsection*{Morphoelastic shell theory} To describe \emph{Volvox} morphogenesis in terms of the axisymmetric deformations of an elastic shell with varying intrinsic stretches and curvatures, we use the shell theory derived in the biologically relevant limit of ``large bending deformations'' in Ref.~\cite{haas21}: For an axisymmetric elastic shell of relative thickness~$h$, we denote by $s$ and $S$ the respective arclengths of the undeformed and deformed cross-sections of the axisymmetric shell. These cross-sections are described by their distances $\rho,r$ from the axis of symmetry and and the tangent angle $\psi$ of the deformed cross-section. The meridional and circumferential stretches and curvatures of the deformed shell are thus
\begin{align}
&f_s(s)=\dfrac{\mathrm{d}S}{\mathrm{d}s},\; f_\phi(s)=\dfrac{r(s)}{\rho(s)},\quad\kappa_s(s)=\dfrac{1}{f_s}\dfrac{\mathrm{d}\psi}{\mathrm{d}s},\;\kappa_\phi(s)=\dfrac{\sin{\psi(s)}}{r(s)},
\end{align}
and we denote by $f_s^0(s),f_{\smash\phi}^0(s),\kappa_s^0(s),\kappa_{\smash\phi}^0(s)$ their intrinsic, preferred values. The differences between the actual stretches and curvatures and their intrinsic values define the shell strains
\begin{subequations}
\begin{align}
&E_s=\dfrac{f_s-f_s^0}{f_s^0},&&E_\phi=\dfrac{f_\phi-f_\phi^0}{f_\phi^0}
\end{align}
and curvature strains
\begin{align}
K_s=\dfrac{\tilde{f_s}\kappa_s-f_s^0\kappa_s^0}{(f_s^0)^2f_\phi^0},&&K_\phi=\dfrac{f_\phi\kappa_\phi-f_\phi^0\kappa_\phi^0}{f_s^0(f_\phi^0)^2} 
\end{align}
\end{subequations}
that appear in the elastic energy
\begin{subequations}\label{eq:energy}
\begin{align}
\mathcal{E}=2\pi\int_{\mathcal{C}}{e(s)\rho(s)\,\mathrm{d}s},
\end{align}
where the integration is along the cross-section $\mathcal{C}$ of the shell, and the energy density is
\begin{align}
e&=\dfrac{C}{2}\left\{h\left[\alpha_{ss}E_s^2+(\alpha_{s\phi}+\alpha_{\phi s})E_sE_\phi+\alpha_{\phi\phi}E_\phi^2\right]\right.\nonumber\\
&\qquad\quad+h^2\left(\beta_{ss}E_sK_s+\beta_{s\phi}E_sK_\phi+\beta_{\phi s}E_\phi K_s+\beta_{\phi\phi}E_\phi K_\phi\right)\nonumber\\
&\qquad\quad+\left.h^3\left[\gamma_{ss}K_s^2+(\gamma_{s\phi}+\gamma_{\phi s})K_sK_\phi+\gamma_{\phi\phi}K_\phi^2\right]\right\},
\end{align}
\end{subequations}
in which $C$ is a material parameter and $\alpha_{ss},\alpha_{s\smash\phi},\alpha_{\smash\phi s},\alpha_{\smash\phi\smash\phi},\beta_{ss},\beta_{s\smash\phi}$, $\beta_{\smash\phi s},\beta_{\smash\phi\smash\phi},\gamma_{ss},\gamma_{s\smash\phi},\gamma_{\smash\phi s},\gamma_{\smash\phi\smash\phi}$ are functions of $\eta=\kappa_s^0h/2f_s^0f_{\smash\phi}^0$ only, the (complicated) explicit expressions for which are given in Eqs.~(58) and (64) of Ref.~\cite{haas21}.

To find the deformed shape of the shell resulting from the imposed intrinsic stretches and curvatures numerically, we solve the boundary value problem associated with Eqs.~\neqref{eq:energy} derived in Ref.~\cite{haas21} using the \texttt{bvp4c} solver of \textsc{Matlab}.
\subsection*{Numerical solution of the toy problems} We solve these equations for a circular cross-section $\rho(s)=\sin{s}$ for \mbox{$\Sigma\leqslant s\leqslant\pi$}. At $s=\pi$, we impose the boundary conditions $r=\smash{\psi}=0$; if $\Sigma>0$, we impose no-force and no-torque conditions at $s=\Sigma$ which, from Ref.~\cite{haas21}, take the form $\alpha_{ss}E_s+\alpha_{s\smash\phi}E_\phi+h(\beta_{ss}K_s+\beta_{s\smash\phi}K_\phi)=0$, $\beta_{ss}E_s+\beta_{\smash\phi s}E_\phi+h(\gamma_{ss}K_s+\gamma_{s\smash\phi}K_\phi)=0$; if $\Sigma=0$, the boundary conditions at $s=0$ are instead $r=\psi=0$.

For toy problem 2, we take $\smash{f_s^0=f_{\smash\phi}^0=1}$, $\smash{\kappa_s^0=\kappa_{\smash\phi}^0}=k$. For toy problem 3, we take $f_s^0=1+f_0=1/f_{\smash\phi}^0$, where $f_0$ is a constant that expresses the anisotropy of contraction, and $\kappa_s^0=\kappa_{\smash\phi}^0=1$. Finally, to describe a possible fast active deformation of the tissue on cutting, we take $f_s^0=f_{\smash\phi}^0=1+F_0(\Sigma+s_0-s)/s_0$ if $\Sigma\leqslant s\leqslant \Sigma+s_0$ for some length scale $s_0$, and $f_s^0=f_{\smash\phi}^0=1$ otherwise, as well as $\kappa_s^0=\kappa_{\smash\phi}^0=1$; results do not seem to depend strongly on the form of the decay of $f_s^0,f_{\smash\phi}^0$ away from $s=\Sigma$ (not shown). The parameter values for \textfigref{fig2}{G--I} are $h=0.3$, $\Sigma=0.3$, $k=0$, $f_0=\pm0.1$, $F_0=0.35$, $s_0=0.2$.
\subsection*{Fitting the morphoelastic model to the experimental data} For a quantitative description of \emph{Volvox} inversion, similarly to Ref.~\cite{haas18a}, we consider again a circular cross-section $\rho(s)=\sin{s}$, but now for $\Sigma\leqslant s\leqslant\pi-P$, where $P$ is the phialopore size. We set $P=0.3$ and $h=0.15$, as in Ref.~\cite{haas18a}. The boundary conditions are those of the toy problems discussed above, except for $s=\pi-P$, where we impose no-force and no-torque conditions.

We define functional forms for the intrinsic stretches and curvature functions $\smash{f_s^0(s),f_\phi^0(s),\kappa_s^0(s),\kappa_\phi^0(s)}$ that allow a minimal representation of the cell shape changes observed during \emph{Volvox} inversion~\cite{hohn11}, described in detail in \SIApp. These functional forms define a number of fitting parameters, which, with $\Sigma=0$, were fitted to the average inversion shapes~\cite{haas18a}, by minimising a fit energy $E_{\mathrm{fit}}$ that measures the difference between the average and fitted shapes using the \textsc{Matlab} function \texttt{fminsearch}, similarly to Ref.~\cite{haas18a}. This fitting is described in detail in \SIApp. 

For numerical ablations, the value of $\Sigma$ was increased from $\Sigma=0$ in these fitted shapes, with boundary conditions as in the toy problems above.
\subsection*{Analytical calculations} Details of the analytical calculations for and additional discussion of the three mechanical toy problems, and the scaling argument for the recoil timescale are given in \SIApp.

}

\showmatmethods{} 

\acknow{The authors are very grateful to Raymond E. Goldstein, in whose research group their collaboration on \emph{Volvox} inversion mechanics started, for many discussions, mentoring, and invariably helpful scientific advice. The authors also thank members of the Cambridge Advanced Imaging Centre, particularly Martin Lenz and Kevin O'Holleran, for their advice and support. The authors gratefully acknowledge funding from the Max Planck Society (P.A.H.), and the Wellcome Trust and the John Templeton Foundation (S.S.M.H.H.).}

\showacknow{} 

\bibliography{main}

\end{document}


\renewcommand{\theequation}{S\arabic{equation}}

\SItext\thispagestyle{firststyle}

This Supplementary Text is divided into four parts. The first three parts provide detailed calculations for the ``toy problems'' in the main text: We first analyse the (small) deformations of a spherical elastic shell with a homogeneous curvature mismatch. We then discuss the (larger) deformations resulting from a circular cut in such a shell, developing the theory and scaling arguments for laser ablations in curved tissues. We then discuss cell sheet curling due to inhomogeneous contraction rather than curvature mismatch. The final part describes the fitting of the elastic model to the averaged inversion shapes in detail.

\section*{Toy Problem 1: deformations of a spherical elastic shell with mismatched intrinsic curvatures}
As in the main text, we consider a thin incompressible elastic shell of undeformed (nondimensional) unit radius, and thickness $h\ll 1$. The shell has uniform intrinsic curvatures $\kappa^0=k$. If $k\not=1$, these are different from the curvatures of the undeformed shell. Due to this curvature mismatch, the shell deforms. We assume the shell to remain spherical and that $k=O(1)$, and denote by $f$ its deformed radius.
\subsection*{Geometry of the deformed shell}
The (spherical) polar radius of a point in the undeformed configuration of the shell is $r=a+\zeta$, where $a$ is the midsurface radius, to be determined in the calculation, and $\zeta$ is a transverse coordinate, with $-h^-\leq\zeta\leq h^+$. By definition, $h^++h^-=h$, and $[(a+h^+)+(a-h^-)]/2=1$, i.e. $a=1-(h^+-h^-)/2$. Analogously, a point in the deformed configuration of the shell has polar radius $\tilde{r}=\tilde{a}+\tilde{\zeta}$, where $\tilde{a}$ and $\tilde{\zeta}$ are the deformed midsurface radius and transverse coordinate, now with $-\tilde{h}^-\leq\tilde{\zeta}\leq\tilde{h}^+$ and hence $\tilde{a}=f-(\tilde{h}^+-\tilde{h}^-)/2$. With respect to spherical polars $(r,\theta,\phi)$, the deformation gradient is diagonal, with principal stretches
\begin{align}
\lambda_r&=\dfrac{\partial\tilde{\zeta}}{\partial\zeta},&&\lambda_\theta=\lambda_\phi=\dfrac{\tilde{a}}{a}\dfrac{1+\tilde{\zeta}/\tilde{a}}{1+\zeta/a}.\label{eq:stretches}
\end{align}
Incompressibility requires $\lambda_r\lambda_\theta\lambda_\phi=1$, which is a differential equation for $\tilde{\zeta}$ as a function of $\zeta$. Imposing $\tilde{\zeta}=0$ at $\zeta=0$, required by the definition of the midsurface, we find its implicit solution
\begin{align}
\left(\dfrac{\tilde{a}}{a}\right)^2\left(\tilde{\zeta}+\dfrac{\tilde{\zeta}^2}{\tilde{a}}+\dfrac{\tilde{\zeta}^3}{3\tilde{a}^2}\right)=\zeta+\dfrac{\zeta^2}{a}+\dfrac{\zeta^3}{3a^2}.\label{eq:cond1}
\end{align}
We now make use of the framework of morphoelasticity~\cite{goriely}: By analogy with \eqsref{eq:stretches}, we define the intrinsic configuration of the cell sheet (which need not be embeddable into three-dimensional space) by its intrinsic stretches
\begin{align}
\lambda_r^0&=\dfrac{\partial Z}{\partial\zeta},&&\lambda_\theta^0=\lambda_\phi^0=\dfrac{1+kZ}{1+\zeta/a}.
\end{align}
To write down these expressions, we have identified $\tilde{a}/a$ and $1/\tilde{a}$ in \eqsref{eq:stretches} as the stretch and curvature of the deformed configuration, respectively, and have replaced them with their respective intrinsic values $1$ and $k$. We will refer to $Z$ as the intrinsic transverse coordinate. We are left to define the midsurfaces (a different choice of these would change the interpretation of $k$, so the existence of this freedom of choice is not surprising). We do so by imposing ${-h^0/2\leq Z\leq h^0/2}$; the intrinsic thickness $h^0$ is determined by the incompressibility condition $\lambda_r^0\lambda_\theta^0\lambda_\phi^0=1$, which integrates to
\begin{align}
\zeta+\dfrac{\zeta^2}{a}+\dfrac{\zeta^3}{3a^2}=Z+kZ^2+\dfrac{k^2Z^3}{3}\label{eq:cond2}
\end{align}
on requiring $Z=0$ at $\zeta=0$. Imposing $Z=\pm h^0/2\Longleftrightarrow\zeta=\pm h^{\pm}\Longleftrightarrow \tilde{\zeta}=\pm\tilde{h}^\pm$ in \eqsref{eq:cond1} and \neqref{eq:cond2} yields
\begin{subequations}\label{eq:heqs}
\begin{align}
\left(\dfrac{f-\dfrac{\tilde{h}^+-\tilde{h}^-}{2}}{1-\dfrac{h^+-h^-}{2}}\right)^2\left[\pm\tilde{h}^\pm+\dfrac{\bigl(\tilde{h}^\pm\bigr)^2}{f-\dfrac{\tilde{h}^+-\tilde{h}^-}{2}}\pm\dfrac{\bigl(\tilde{h}^\pm\bigr)^3}{3\left(f-\dfrac{\tilde{h}^+-\tilde{h}^-}{2}\right)^2}\right]&=\pm h^\pm+\dfrac{(h^\pm)^2}{1-\dfrac{h^+-h^-}{2}}\pm\dfrac{(h^\pm)^3}{3\left(1-\dfrac{h^+-h^-}{2}\right)^2}\\&=\pm\dfrac{h^0}{2}+\dfrac{k\bigl(h^0\bigr)^2}{4}\pm\dfrac{k^2\bigl(h^0\bigr)^3}{24}.
\end{align}
\end{subequations}
Together with the condition $h^++h^-=h$, these form a system of equations for $h^\pm$, $\tilde{h}^\pm$, and $h^0$, the solution of which defines the deformed geometry of the shell.

\subsection*{Elasticity of the deformed shell} Let $\tens{F}$ and $\tens{F}^{\mathbfsf{0}}$ be the diagonal tensors with principal stretches $\lambda_r,\lambda_\theta,\lambda_\phi$ and $\lambda_r^0,\lambda_\theta^0,\lambda_\phi^0$, respectively, relative to the standard basis of spherical polars. By the fundamental relation of morphoelasticity~\cite{goriely}, the elastic deformation gradient is $\smash{\tens{F}\bigl(\tens{F}^{\mathbfsf{0}}\bigr)^{-1}}$. The elastic deformation gradient therefore has principal stretches
\begin{align}
\Lambda_\theta=\Lambda_\phi&=\dfrac{\lambda_\theta}{\lambda_\theta^0}=\dfrac{\lambda_\phi}{\lambda_\phi^0}=\dfrac{\tilde{a}}{a}\dfrac{1+\tilde{\zeta}/\tilde{a}}{1+kZ}\equiv\Lambda,&\Lambda_r&=\dfrac{\lambda_r}{\lambda_r^0}=\dfrac{1/\lambda_\theta\lambda_\phi}{1/\lambda_\theta^0\lambda_\phi^0}=\dfrac{1}{\Lambda^2},
\end{align}
where we have used the incompressibility condition. The first two principal invariants of the associated Cauchy--Green tensor are thus $\mathcal{I}_1=2\Lambda^2+\Lambda^{-4}$ and $\mathcal{I}_2=\Lambda^4+2\Lambda^{-2}$.

Since the problem that we are solving is \emph{not} a leading-order problem (in an expansion in the small thickness of the shell), the constitutive assumptions matter. We therefore consider general elastic energy densities with a regular power series expansion~\cite{dervaux09}, viz.,
\begin{align}
e&=\dfrac{1}{2}\sum_{m=0}^\infty{\sum_{n=0}^\infty{C_{mn}(\mathcal{I}_1-3)^m(\mathcal{I}_2-3)^n}},
\end{align}
where we have set $C_{00}=0$ without loss of generality, and we may assume that $C_{01}+C_{10}>0$ for the material to have a positive bulk modulus~\cite{haas21}. By incompressibility, we have equality $\mathrm{d}V^0=\mathrm{d}V$ of the volume elements of the intrinsic and undeformed configurations. Explicitly,
\begin{align}
\mathrm{d}V^0=\mathrm{d}V&=(a+\zeta)^2\sin{\theta}\,\mathrm{d}\zeta\,\mathrm{d}\theta\,\mathrm{d}\phi=h a^2(1+khz)^2\sin{\theta}\,\mathrm{d}z\,\mathrm{d}\theta\,\mathrm{d}\phi,
\end{align}
using incompressibility again, and where $z=Z/h$. The elastic energy of the shell is the integral of the energy density with respect to the intrinsic configuration, viz.,
\begin{align}
\mathcal{E}=\int_{\mathcal{V}^0}{e\,\mathrm{d}V^0}=4\pi h\int_{-h^0/2h}^{h^0/2h}{\hspace{-2mm}ea^2(1+khz)^2\,\mathrm{d}z}.
\end{align}

\subsection*{Asymptotic solution} We now determine the deformed radius $f$ by minimising $\mathcal{E}$ by asymptotic expansion for $h\ll 1$. To keep the algebraic expressions that arise in the calculation (somewhat) simple, it will be useful to divide the asymptotic solution into two steps.
\subsubsection*{Leading-order calculation} We begin by showing that $f=1+O(h^2)$, recovering the scaling behaviour obtained in the main text using a shell theory. To this end, we introduce expansions
\begin{align}
h^\pm&=\dfrac{h}{2}+h^2h^\pm_2+O(h^3),&\tilde{h}^\pm&=\dfrac{h}{2}+h^2\tilde{h}^\pm_2+O(h^3), &h^0&=h+h^2h^0_2+O(h^3),& f&=1+hf_1+h^2f_2+O(h^3).\label{eq:leadingexp}
\end{align}
Using \textsc{Mathematica} (Wolfram, Inc.), \eqsref{eq:heqs} yield
\begin{align}
&h^+_2=-h^-_2=\dfrac{k-1}{4}, &&\tilde{h}^+_2=\dfrac{k-1}{4}-f_1,&&\tilde{h}^-_2=-\dfrac{k-1}{4}-f_1,&&h^0_2=0.\label{eq:lo1}
\end{align}
We then set $Z=hz$, $\tilde{\zeta}=h\tilde{\zeta}_1+h^2\tilde{\zeta}_2+O(h^3)$ and solve \eqsref{eq:cond1} and \neqref{eq:cond2} for $\tilde{\zeta}_1,\tilde{\zeta}_2$ as functions of $z$. We find
\begin{align}
&\tilde{\zeta}_1=z,&&\tilde{\zeta}_2=-2f_1z+(k-1)z^2,\label{eq:lo2}
\end{align}
and hence compute
\begin{align}
\mathcal{E}=2\pi(C_{10}+C_{01})\bigl[12f_1^2+(k-1)^2\bigr]h^2+O(h^3),
\end{align}
which is minimised for all $k$ at $f_1=0$. This proves that $f=1+O(h^2)$ as claimed.

\subsubsection*{Next-order calculation} To determine the leading-order deformation of the shell, we extend expansions~\neqref{eq:leadingexp} using the leading-order results~\neqref{eq:lo1}, writing
\begin{subequations}
\begin{align}
h^\pm&=\dfrac{h}{2}\pm \dfrac{h^2(k-1)}{4}+h^3h^\pm_3+h^4h^\pm_4+O(h^5),&\tilde{h}^\pm&=\dfrac{h}{2}\pm \dfrac{h^2(k-1)}{4}+h^3\tilde{h}^\pm_3+h^4\tilde{h}^\pm_4+O(h^5),
\end{align}
and
\begin{align}
h^0&=h+h^3h^0_3+h^4h^0_4+O(h^5),&f&=1+h^2f_2+h^3f_3+h^4f_4+O(h^5).
\end{align}
\end{subequations}
Solving \eqsref{eq:heqs} order-by-order for $h^\pm_3,h^\pm_4,\tilde{h}^\pm_3,\tilde{h}^\pm_4,h^0_3,h^0_4$, leaving only $f_2,f_3,f_4$ as unknowns, we find
\begin{subequations}
\begin{align}
h^\pm_3&=0,&\tilde{h}^\pm_3&=-f_2,&h^0_3&=-\dfrac{1}{12}\left(5-6k+k^2\right)
\end{align}
and
\begin{align}
h^\pm_4&=\pm\dfrac{1}{48}\left(3-10k+9k^2-2k^3\right),&\tilde{h}^\pm_4&=\pm\dfrac{1}{48}\left(3-10k+9k^2-2k^3\right)-f_3\pm\dfrac{f_2}{4}(5-2k),&h^0_4&=0.
\end{align}
\end{subequations}
We set $Z=hz$ again and write $\tilde{\zeta}=hz+(k-1)h^2z^2+\tilde{\zeta}_3h^3+\tilde{\zeta}_4h^4+O(h^5)$ using the leading-order results~\neqref{eq:lo2}. From \eqsref{eq:cond1} and \neqref{eq:cond2}, we obtain 
\begin{align}
\tilde{\zeta}_3&=-2f_2z+\dfrac{z^3}{3}\left(5-6k+k^2\right),&\tilde{\zeta}_4&=-2f_3z-\dfrac{z^2}{4}\left[k-1-4(5-2k)f_2\right]-\dfrac{z^4}{3}\left(10-15k+9k^2\right).
\end{align}
Using \textsc{Mathematica}, we compute
\begin{align}
\dfrac{\mathcal{E}}{2\pi}&= (C_{10}+C_{01})(k-1)^2h^2\nonumber\\
&\qquad+\left\{12(C_{10}+C_{01})f_2^2-(k-1)[C_{10}(5k-11)+C_{01}(k-7)]f_2+g(k;C_{10},C_{01},C_{20},C_{11},C_{02})\right\}h^4+O(h^5),
\end{align}
where $g$ is a quartic polynomial in $k$ with coefficients depending on the material parameters $C_{10},C_{01},C_{20},C_{11},C_{02}$, the explicit expression of which is of no relevance to our calculation. Minimising $\mathcal{E}$ to determine $f_2$ finally yields
\begin{align}
f&=1-\dfrac{h^2}{12}(k-1)\dfrac{(11C_{10}+7C_{01})-(5C_{10}+C_{01})k}{2(C_{10}+C_{01})}+O(h^3).  \label{eq:f}
\end{align}
\subsubsection*{Surprising mechanical behaviour}While this result has the same scaling behaviour as the result derived in the main text from a shell theory not formally valid for these small deformations, it features a complicated prefactor depending on the material parameters $C_{10},C_{01}$ and also on $k$ that is not predicted by the shell theory, but that gives rise to surprising and somewhat counterintuitive behaviour: indeed, the $O(h^2)$ correction in \eqref{eq:f} vanishes and changes sign not only at $k=1$ as expected, but also at $k=k_\ast$, where $k_\ast=(11C_{10}+7C_{01})/(5C_{10}+C_{01})=1+6(C_{10}+C_{01})/(5C_{10}+C_{01})$. From $C_{10}+C_{01}>0$, as required for the material to have a positive bulk modulus, it follows that $k_\ast\gtrless 1$ iff $5C_{10}+C_{01}\gtrless 0$, and hence $f\gtrless 1$ iff $1\lessgtr k\lessgtr k_\ast$ for $5C_{10}+C_{01}\gtrless 0$. As expected, this shows that the sphere grows (shrinks) if $k$ is just below (above) unit value, but it is surprising that the sphere grows if $k>k_\ast\;(>1)$ if $5C_{10}+C_{01}>0$ and shrinks if $k<k_\ast\;(<1)$ if $5C_{10}+C_{01}<0$. It is likely that the reason for which the intuition that the sphere should simply grow and shrink if $k<1$ and $k>1$ respectively fails because these are very small deformations beyond leading order, asymptotically smaller than the thickness of the shell.

In this context, we also note that (again because the problem is not a leading-order problem), the definition of the midsurfaces (and hence of the intrinsic curvatures) matters: Had we chosen to define the midsurface in the undeformed configuration rather than in the intrinsic configuration, we would have obtained a different answer, simply because the middle surface of the undeformed configuration does not map to the middle surface of the intrinsic configuration, and so the intrinsic curvatures do not directly correspond to each other. This ambivalence does not however change the fact that there is ``surprising'' behaviour as discussed in the previous paragraph.

\section*{Toy Problem 2: circular ablation of a spherical elastic shell with mismatched intrinsic curvatures}
As in the main text, we consider a circular cut of angular extent $2\Sigma$ in a thin incompressible elastic spherical shell of unit radius, of thickness $h\ll 1$, and with intrinsic curvatures $\kappa^0=k$. If $k\not=1$, these are mismatched; to determine the resulting deformation of the shell, we take cylindrical polar coordinates defined by the axis of the cut.

\subsection*{Governing equations and boundary conditions} With respect to these coordinates, the undeformed spherical shell has polar radius $\rho(s)=\sin{s}$, where $s$ is arclength, measured from the centre of the cut, so its rim corresponds to $s=\Sigma$. The deformed shell has arclength $S(s)$, corresponding polar radius $r(s)$, and tangent angle $\psi(s)$. The stretches and curvatures of the deformed shell are thus
\begin{align}
&f_s(s)=\dfrac{\mathrm{d}S}{\mathrm{d}s},&&f_\phi(s)=\dfrac{r(s)}{\rho(s)},&&\kappa_s(s)=\dfrac{1}{f_s}\dfrac{\mathrm{d}\psi}{\mathrm{d}s},&& \kappa_\phi(s)=\dfrac{\sin{\psi(s)}}{r(s)}.
\end{align}
These define the shell strains and curvature strains
\begin{align}
E_s&=f_s-1,&E_\phi&=f_\phi-1, &K_s&=f_s\kappa_s-k,&K_\phi&=f_\phi\kappa_\phi-k. \label{eq:str}
\end{align}
For this toy problem, the curvatures remain small compared to the shell thickness. We do not therefore need to use the ``large bending theory''~\cite{haas21} used in the numerical calculations in the main text for our asymptotic calculations. On defining
\begin{align}
N_s&=2E_s+E_\phi,&N_\phi&=E_s+2E_\phi,&
M_s&=2K_s+K_\phi,&M_\phi&=K_s+2K_\phi,\label{eq:stresses}
\end{align}
and letting $\varepsilon=h/\sqrt{12}$, the deformed configuration of the shell satisfies the (nondimensionalised) force-balance equations~\cite{haas15}
\begin{subequations}\label{eq:goveq0}
\begin{align}
\varepsilon^2\dfrac{\mathrm{d}}{\mathrm{d}s}(rM_s)&=f_s\left(rN_s\tan{\psi}+\varepsilon^2M_\phi\cos{\psi}\right),&\dfrac{\mathrm{d}}{\mathrm{d}s}(rN_s\sec{\psi})&=f_sN_\phi,
\end{align}
complemented by the geometric equations
\begin{align}
&\dfrac{\mathrm{d}r}{\mathrm{d}s}=f_s\cos{\psi},&& \dfrac{\mathrm{d}\psi}{\mathrm{d}s}=f_s\kappa_s.\label{eq:drpsi}
\end{align}
\end{subequations}
It will however turn out to be convenient to express these equations in terms of $\psi$, $E_s$, $E_\phi$, $N_s$ for the asymptotic solution. To this end, rearranging the first of~\eqsref{eq:drpsi} using definitions~\neqref{eq:str}, we notice that
\begin{align}
\dfrac{\mathrm{d}}{\mathrm{d}s}\bigl(E_\phi\rho(s)\bigr)=f_s\cos{\psi}-\rho'(s). \label{eq:dEphi0}
\end{align}
Moreover, we obtain, from \eqsref{eq:str} and~\neqref{eq:stresses}, the algebraic relations
\begin{align}
&2N_\phi=3E_\phi+N_s,&& 2f_s=2+N_s-E_\phi.
\end{align}
Inserting these into \eqsref{eq:goveq0} and \neqref{eq:dEphi0} and expanding, we conclude that deformed configuration of the shell is described by the equations\begin{subequations}\label{eq:goveq}
\begin{align}
&4\dfrac{\mathrm{d}}{\mathrm{d}s}\left[(1+E_\phi)N_s\sec{\psi}\sin{s}\right]=(2+N_s-E_\phi)(3E_\phi+N_s),\label{eq:goveqa}\\
&2\varepsilon^2\dfrac{\mathrm{d}}{\mathrm{d}s}\left[(1+E_\phi)\sin{s}\left(2\dfrac{\mathrm{d}\psi}{\mathrm{d}s}+\dfrac{\sin{\psi}}{\sin{s}}-3k\right)\right]=(2+N_s-E_\phi)\left[(1+E_\phi)N_s\tan{\psi}\sin{s}+\varepsilon^2\cos{\psi}\left(\dfrac{\mathrm{d}\psi}{\mathrm{d}s}+2\dfrac{\sin{\psi}}{\sin{s}}-3k\right)\right],\label{eq:goveqb}\\
&2\dfrac{\mathrm{d}}{\mathrm{d}s}\left(E_\phi\sin{s}\right)=2(\cos{\psi}-\cos{s})+(N_s-E_\phi)\cos{\psi}.\label{eq:dEphi} 
\end{align}
\end{subequations}
The boundary conditions at the rim of the cut, $s=\Sigma$, are
\begin{align}
&N_s(\Sigma)=0,&&M_s(\Sigma)=0\quad\Longleftrightarrow\quad 2\psi'(\Sigma)+\dfrac{\sin{\psi(\Sigma)}}{\sin{\Sigma}}=3k.\label{eq:bcs}
\end{align}
\subsection*{Asymptotic solution} In the asymptotic limit $\varepsilon\ll 1$, the shell deforms in a region of characteristic extent $\delta\ll 1$ near $s=\Sigma$. In this inner region, we let $s=\Sigma+\delta\xi$, defining an inner coordinate $\xi$. Let $\beta(\xi)=\psi(s)-s$ be the deviation of the tangent angle due to the cut; this deviation is small, $\beta\ll 1$. Expanding the no-torque boundary condition,
\begin{align}
\dfrac{2}{\delta}\dot{\beta}(0)+\beta(0)\cot{\Sigma}=3(k-1), \label{eq:torquebc}
\end{align}
where the dot denotes a derivative with respect to $\xi$. This condition therefore requires $\beta=O(\delta)$. Scaling from~\eqsref{eq:goveq}, we also find
\begin{align}
&\dfrac{N_s}{\delta}\sim E_\phi,&&\dfrac{\varepsilon^2\beta}{\delta^2}\sim N_s,&&\dfrac{E_\phi}{\delta}\sim\max{\{\beta,N_s,E_\phi\}}. 
\end{align}
The first scaling yields $N_s\ll E_\phi$, and hence the third scaling becomes $E_\phi/\delta\sim\beta$. The scalings then reduce to $\delta\sim\smash{\sqrt{\varepsilon}}$, and hence $E_\phi\sim\smash{\delta^2}$, $N_s\sim\smash{\delta^3}$. [The asymptotic expansion of elasticity~\cite{haas21} that leads to the shell equations~\neqref{eq:goveq0} includes the scalings $E_s,E_\phi=O(\varepsilon)$ and $K_s,K_\phi=O(1)$, consistent with these scalings.] We therefore posit regular expansions
\begin{align}
\psi&=s+\delta\left[b+O(\delta)\right],&
E_\phi&=\delta^2\left[e+O(\delta)\right],&
N_s&=\delta^3\cot{\Sigma}\left[n+O(\delta)\right],\label{eq:scalings}
\end{align}
so that, on Taylor expanding and at leading order, \eqsref{eq:goveq} become 
\begin{align}
&2\dot{n}=3e,&&2\ddot{b}=n,&&\dot{e}=-b.\label{eq:diffeqs}
\end{align}
The leading-order problem is therefore, on recalling~\eqsref{eq:bcs} and \neqref{eq:torquebc},
\begin{align}
4\ddddot{b}+3b=0\qquad\text{subject to}\quad b\rightarrow 0\text{ as }\xi\rightarrow\infty,\;2\dot{b}(0)=3(k-1),\;\ddot{b}(0)=0.
\end{align}
Its solution is
\begin{align}
b(\xi)=-3^{3/4}(k-1)\mathrm{e}^{-3^{1/4}\xi/2}\cos{\left(\dfrac{3^{1/4}\xi}{2}\right)}.\label{eq:bsol}
\end{align}
\subsubsection*{Cut opening} One experimentally accessible geometric measure that characterises the deformation resulting from the curvature mismatch is the cut opening $o$, i.e. the amount by which the cut opens during the recoil following the ablation. To compute $o$, we note that the radial displacement $u(s)=r(s)-\rho(s)$ of the shell satisfies the differential equation
\begin{align}
\dfrac{\mathrm{d}u}{\mathrm{d}s}=f_s\cos{\psi}-\cos{s}\quad\Longrightarrow\quad \dfrac{\mathrm{d}u}{\mathrm{d}\xi}=-\delta^2b\sin{\Sigma}+O(\delta^3),
\end{align}
were we have made use of $f_s=1+O(\delta^2)$. Now $o=2[u(\xi=0)-u(\xi\rightarrow\infty)]$, twice the radial displacement of the cut edge across the boundary layer. We thus obtain
\begin{align}
o=2\delta^2\sin{\Sigma}\int_0^\infty{\!b(\xi)\,\mathrm{d}\xi}+O(\delta^3)=-h(k-1)\sin{\Sigma}+O(\delta^3).\label{eq:Usol}
\end{align}
As expected, this shows that the rim of the cut curls outwards if $k<1$, and inwards if $k>1$. We also note that $o$ is proportional to the radius $\sin{\Sigma}$ of the ablation.
\subsubsection*{Cut rotation and cut displacement: breakdown of the leading-order asymptotics} There are (at least) two other exerimentally accessible measures of the deformation: the rotation $\varrho$ of the edge of the cut, and its displacement $d$ during the recoil. By definition,
\begin{align}
\varrho=\delta b(0)+O(\delta^2)=-\sqrt{\dfrac{3h}{2}}(k-1)+O(\delta^2). \label{eq:cutr}
\end{align}
Moreover, the calculation of the cut opening shows that the radial displacement of the cut edge is $-(h/2)(k-1)\sin{\Sigma}+O(\delta^3)$. Similarly, the axial displacement of the cut edge is $-(h/2)(k-1)\cos{\Sigma}+O(\delta^3)$, whence
\begin{align}
d=\dfrac{h}{2}|k-1|+O(\delta^3). \label{eq:cutd}
\end{align}
Both leading-order results are independent of the cut size $\Sigma$. This is clearly absurd, because we must have $\varrho,d\rightarrow 0$ as $\Sigma\rightarrow 0$. This signals a breakdown of the leading-order asymptotics that we will analyse in the next subsection.
\subsection*{Asymptotic solution for small ablations} We now address the breakdown of the leading-order asymptotic solution for small cut sizes that we highlighted above when deriving the leading-order expressions for the cut rotation and cut displacement.
\subsubsection*{Breakdown of asymptoticity} To understand this breakdown of the asymptotic validity of the leading-order solution, we expand further, writing
\begin{align}
\psi&=s+\delta\left[b_0+\delta b_1+O(\delta^2)\right],&
E_\phi&=\delta^2\left[e_0+\delta e_1+O(\delta^2)\right],&
N_s&=\delta^3\cot{\Sigma}\left[n_0+\delta n_1+O(\delta^2)\right], 
\end{align}
where $b_0$ is given by \eqref{eq:bsol}, and  $n_0=2\ddot{b}_0$, $e_0=4\dddot{b}_0/3$ from \eqsref{eq:diffeqs}. On expanding \eqsref{eq:goveq} and the boundary conditions pertaining thereto using \textsc{Mathematica}, we obtain
\begin{align}
4\ddddot{b}_1+3b_1=\cot{\Sigma}\left(4\ddot{b}_0^2+8\dot{b}_0\dddot{b}_0-8\dddot{b}_0-\tfrac{9}{2}b_0^2\right)\quad\text{subject to }\dot{b}_1(0)=-\tfrac{1}{2}b_0(0)\cot{\Sigma},\;\ddot{b}_1(0)=-\dot{b}_0(0)\cot{\Sigma}.\label{eq:Odelta}
\end{align}
This shows that $b_1(\xi)$ is proportional to $\cot{\Sigma}$, so the expansion $b=b_0+\delta b_1+O(\delta^2)$ loses asymptoticity when $\delta\cot{\Sigma}=O(1)$, which is when $\Sigma=O(\delta)$.
\subsubsection*{Asymptotic scalings for small ablations} We must therefore derive the asymptotic scalings for small cut sizes $\Sigma=O(\delta)$. We thus write $\Sigma=\delta\sigma$, where $\sigma=O(1)$. The scaling $\beta=O(\delta)$ still holds, but establishing the other scalings takes more effort. 

First, expanding \eqref{eq:goveqa} shows that $E_\phi\lesssim N_s$. If $E_\phi\ll N_s$, then we may posit $N_s=\nu n+o(\nu)$, $E_\phi=o(\nu)$ for some $\nu\ll 1$, and \eqref{eq:goveqa} becomes, at leading order,
\begin{align}
2\dfrac{\mathrm{d}}{\mathrm{d}\xi}\left[(\sigma+\xi)n\right]=n\quad\Longrightarrow\quad n(\xi)=\dfrac{C_1}{\sqrt{\xi+\sigma}},
\end{align}
where $C_1$ is a constant of integration. The no-force condition at the rim of the cut is $n(0)=0$, yielding $C_1=0$ and hence $n\equiv 0$, which is a contradiction. This shows that $N_s\sim E_\phi$.

Next, on expanding \eqref{eq:dEphi}, we find that $\delta^2\lesssim \smash{E_\phi}\sim N_s$. If $\smash{\delta^2}\ll E_\phi\sim N_s$, then $N_s=\nu n+o(\nu)$, $E_\phi=\nu e+o(\nu)$, for some $\nu\ll 1$, and hence, from \eqsref{eq:goveqa} and \neqref{eq:dEphi},
\begin{subequations}
\begin{align}
2\dfrac{\mathrm{d}}{\mathrm{d}\xi}\left[(\sigma+\xi)n\right]=3e+n,\;
2\dfrac{\mathrm{d}}{\mathrm{d}\xi}\left[(\sigma+\xi)e\right]=n-e\quad\Longrightarrow\quad 2(\sigma+\xi)\dot{n}=3e-n,\;
2(\sigma+\xi)\dot{e}=n-3e.
\end{align}
Summing these equations yields $n+e=\text{const.}$, and then integrating the first gives, in particular,
\begin{align}
n(\xi)=C_2+\dfrac{C_3}{(\sigma+\xi)^2}, 
\end{align}
\end{subequations}
wherein $C_2,C_3$ are more constants of integration. The no-force condition at the rim of the cut is $n(0)=0$, while matching to the undeformed configuration $f_s=f_\phi=1$ as $\xi\rightarrow\infty$ requires $n\rightarrow0$ in this limit. Hence $C_2=C_3=0$, so $n\equiv0$. This is another contradiction, so $E_\phi\sim N_s\sim\delta^2$.

Finally, expanding \eqref{eq:goveqb} gives $N_s\delta^2\lesssim\varepsilon^2$. If $N_s\delta^2\ll\varepsilon^2$, then, on positing $\psi=s+\delta b+O(\delta^2)$ again, \eqref{eq:goveqb} becomes, at leading order,
\begin{subequations}
\begin{align}
&\dfrac{\mathrm{d}}{\mathrm{d}s}\left[2(\sigma+\xi)\dot{b}+b\right]=\dot{b}+\dfrac{2b}{\sigma+\xi}\quad\Longrightarrow\quad(\xi+\sigma)^2\ddot{b}+(\xi+\sigma)\dot{b}-b=0,
\end{align}
which is a homogeneous equation with general solution
\begin{align}
b(\xi)=C_4(\xi+\sigma)+\dfrac{C_5}{\xi+\sigma},
\end{align}
\end{subequations}
where $C_4,C_5$ are yet more constants of integrations. Matching to the undeformed shell as $\xi\rightarrow\infty$ requires $\beta\rightarrow0$ in this limit, and hence $C_4=0$. Let $v(s)$ be the vertical displacement of the shell, which satisfies $\mathrm{d}v/\mathrm{d}s=f_s\sin{\psi}-\sin{s}$. Expanding yields $\dot{v}=\delta^2C_5(\xi+\sigma)^{-1}+O(\delta^3)$, which does not integrate to a finite displacement unless $C_5=0$, i.e. unless $b\equiv 0$. 

This is a final contradiction, so $N_s\delta^2\sim\varepsilon^2$. Since we have shown above that $E_\phi\sim N_s\sim\delta^2$, this yields $\delta\sim\sqrt{\varepsilon}$, which finally determines the asymptotic balance. In particular, $E_s,E_\phi=O(\varepsilon)$, as expected from the scalings of the asymptotic expansion underlying the derivation of the shell theory~\cite{haas21}. 
\subsubsection*{Asymptotic governing equations for small ablations}
Having determined the asymptotic balance for $\Sigma=O(\delta)$, we can now obtain the governing equations in this limit. According to the scalings derived above, we replace \eqsref{eq:scalings} with
\begin{align}
\psi&=s+\delta\left[b+O(\delta)\right],&
E_\phi&=\delta^2\left[e+O(\delta)\right],&
N_s&=\delta^2\left[n+O(\delta)\right].\label{eq:scalings2} 
\end{align}
Equations~\neqref{eq:goveq} yield
\begin{align}
2(\sigma+\xi)\dot{n}&=3e-n,&
2(\sigma+\xi)^2\ddot{b}&=-2(\sigma+\xi)\dot{b}+2b+(\sigma+\xi)^3n+(\sigma+\xi)^2bn,&
2(\sigma+\xi)\dot{e}&=n-3e-2(\sigma+\xi)b-b^2.\label{eq:proba}
\end{align} 
On eliminating $e$, these equations reduce to the pair of nonlinear second-order equations
\begin{align}
4(\sigma+\xi)^2\ddot{n}&=-12(\sigma+\xi)\dot{n}-6(\sigma+\xi)b-3b^2,&
2(\sigma+\xi)^2\ddot{b}&=-2(\sigma+\xi)\dot{b}+2b+(\sigma+\xi)^3n+(\sigma+\xi)^2bn,\label{eq:prob}
\end{align} 
which are subject to the boundary conditions
\begin{align}
&n(0)=0,&&2\dot{b}(0)+\dfrac{b(0)}{\sigma}=3(k-1), &&b,n\rightarrow 0\text{ as }\xi\rightarrow\infty.\label{eq:probbc}
\end{align}
Since \eqsref{eq:prob} are nonlinear, their solution is not proportional to $k-1$, unlike the solution of the leading-order problem for $\Sigma=O(1)$. They must be solved numerically in general.

\subsubsection*{Approximate solution for small ablations} It is therefore all the more remarkable that it is possible to find an analytical approximate solution of \eqsref{eq:prob} that satisfies the boundary conditions~\neqref{eq:probbc} and for which the cut rotation and cut displacement match onto \eqsref{eq:cutr} and \neqref{eq:cutd} in the limit $\sigma\rightarrow\infty$.

Our starting point is the observation that the boundary conditions for $\xi\rightarrow\infty$ imply that $b\ll\sigma+\xi$ there, and hence the nonlinear terms in~\eqsref{eq:prob} can be neglected there. On setting $x=\sigma+\xi$, \eqsref{eq:prob} reduce to
\begin{align}
2x^2\ddot{n}&=-6x\dot{n}-3xb,&2x^2\ddot{b}&=-2x\dot{b}+2b+x^3n,\label{eq:prob2}
\end{align}
in the limit $x\rightarrow\infty$, where dots now denote derivatives with respect to $x$. The solution of these equations subject to boundary conditions \neqref{eq:probbc} will constitute our approximate solution.

To obtain this solution, we introduce the differential operator $\mathscr{D}=x^2\mathrm{d}^2/\mathrm{d}x^2+x\mathrm{d}/\mathrm{d}x-1$. By definition, $\mathscr{D}b=x^2(xn)/2$. Observe that $\mathscr{D}(xn)=x^3\ddot{n}+3x^2\dot{n}=-3x^2b/2$ using \eqsref{eq:prob2}, so
\begin{align}
\mathscr{D}\left(b\pm\dfrac{\mathrm{i}}{\sqrt{3}}xn\right)=\mp\dfrac{\sqrt{3}}{2}\mathrm{i}x^2\left(b\pm\dfrac{\mathrm{i}}{\sqrt{3}}xn\right)=\left(\dfrac{3^{1/4}}{\sqrt{2}}\mathrm{e}^{\mp\mathrm{i}\pi/4}x\right)^2\left(b\pm\dfrac{\mathrm{i}}{\sqrt{3}}xn\right), 
\end{align}
which are complex Bessel equations~\cite{abramowitz}, the solutions of which can be written as
\begin{align}
&b+\dfrac{\mathrm{i}}{\sqrt{3}}xn=c_1\mathcal{K}_1\left(\dfrac{3^{1/4}}{\sqrt{2}}\mathrm{e}^{-\mathrm{i}\pi/4}x\right)+c_2\mathcal{I}_1\left(\dfrac{3^{1/4}}{\sqrt{2}}\mathrm{e}^{-\mathrm{i}\pi/4}x\right),&&b-\dfrac{\mathrm{i}}{\sqrt{3}}xn=c_3\mathcal{K}_1\left(\dfrac{3^{1/4}}{\sqrt{2}}\mathrm{e}^{\mathrm{i}\pi/4}x\right)+c_4\mathcal{I}_1\left(\dfrac{3^{1/4}}{\sqrt{2}}\mathrm{e}^{\mathrm{i}\pi/4}x\right),
\end{align}
where $c_1,c_2,c_3,c_4$ are constants of integration, and where $\mathcal{I}_1$, $\mathcal{K}_1$ are the modified Bessel functions of the first order~\cite{abramowitz} which satisfy $\mathcal{I}_1(z)\sim (2\pi z)^{-1/2}\mathrm{e}^z$ and $\mathcal{K}_1(z)\sim (\pi/2z)^{1/2}\mathrm{e}^{-z}$ as $z\rightarrow\infty$ for $\operatorname{Re}(z)>0$. Accordingly, $b,n\rightarrow 0$ as $x\rightarrow\infty$ requires $c_2=c_4=0$. Hence
\begin{subequations}
\begin{align}
b(x)&=\dfrac{c_1}{2}\mathcal{K}_1\left(\dfrac{3^{1/4}}{\sqrt{2}}\mathrm{e}^{-\mathrm{i}\pi/4}x\right)+\dfrac{c_3}{2}\mathcal{K}_1\left(\dfrac{3^{1/4}}{\sqrt{2}}\mathrm{e}^{\mathrm{i}\pi/4}x\right),&n(x)&=\dfrac{\sqrt{3}}{x}\left[-\dfrac{\mathrm{i}c_1}{2}\mathcal{K}_1\left(\dfrac{3^{1/4}}{\sqrt{2}}\mathrm{e}^{-\mathrm{i}\pi/4}x\right)+\dfrac{\mathrm{i}c_3}{2}\mathcal{K}_1\left(\dfrac{3^{1/4}}{\sqrt{2}}\mathrm{e}^{\mathrm{i}\pi/4}x\right)\right],\\
&=c_1'\operatorname{ker}_1{\left(\dfrac{3^{1/4}x}{\sqrt{2}}\right)} +c_2'\operatorname{kei}_1{\left(\dfrac{3^{1/4}x}{\sqrt{2}}\right)},&
&=\dfrac{\sqrt{3}}{x}\left[-c_1'\operatorname{kei}_1{\left(\dfrac{3^{1/4}x}{\sqrt{2}}\right)} +c_2'\operatorname{ker}_1{\left(\dfrac{3^{1/4}x}{\sqrt{2}}\right)}\right],
\end{align}
\end{subequations}
where $c_1'=\mathrm{i}(c_3-c_1)/2$, $c_2'=-(c_1+c_3)/2$, and where we use $\operatorname{ker}_0,\operatorname{ker}_1,\dots$ and $\operatorname{kei}_0,\operatorname{kei}_1,\dots$ to denote the Kelvin functions~\cite{abramowitz}. Applying the remaining, first two boundary conditions in \eqsref{eq:probbc} using \textsc{Mathematica} and setting $\varsigma=3^{1/4}\sigma/\sqrt{2}$, we obtain
\begin{subequations}
\begin{align}
c'_1&=\dfrac{3^{3/4}\sqrt{2}(1-k)\varsigma \operatorname{ker}_1{\varsigma}}{\left(\operatorname{ker}_1{\varsigma}\right)^2+\left(\operatorname{kei}_1{\varsigma}\right)^2+\sqrt{2}\varsigma\left[\operatorname{ker}_1{\varsigma}\left(\operatorname{ker}_0{\varsigma}+\operatorname{kei}_0{\varsigma}\right)+\operatorname{kei}_1{\varsigma}\left(\operatorname{kei}_0{\varsigma}-\operatorname{ker}_0{\varsigma}\right)\right]},\\
c'_2&=\dfrac{3^{3/4}\sqrt{2}(1-k)\varsigma \operatorname{kei}_1{\varsigma}}{\left(\operatorname{ker}_1{\varsigma}\right)^2+\left(\operatorname{kei}_1{\varsigma}\right)^2+\sqrt{2}\varsigma\left[\operatorname{ker}_1{\varsigma}\left(\operatorname{ker}_0{\varsigma}+\operatorname{kei}_0{\varsigma}\right)+\operatorname{kei}_1{\varsigma}\left(\operatorname{kei}_0{\varsigma}-\operatorname{ker}_0{\varsigma}\right)\right]}.
\end{align}
\end{subequations}
From this solution, we obtain the leading-order rotatation of the edge of the cut
\begin{align}
\varrho\sim \delta b(\sigma)=\dfrac{\sqrt{3}\varsigma\left[\left(\operatorname{ker}_1{\varsigma}\right)^2+\left(\operatorname{kei}_1{\varsigma}\right)^2\right]}{\left(\operatorname{ker}_1{\varsigma}\right)^2+\left(\operatorname{kei}_1{\varsigma}\right)^2+\sqrt{2}\varsigma\left[\operatorname{ker}_1{\varsigma}\left(\operatorname{ker}_0{\varsigma}+\operatorname{kei}_0{\varsigma}\right)+\operatorname{kei}_1{\varsigma}\left(\operatorname{kei}_0{\varsigma}-\operatorname{ker}_0{\varsigma}\right)\right]}\sqrt{h}(1-k). 
\end{align}
In particular, $\varrho\rightarrow 0$ as $\sigma\rightarrow 0$: the cut rotation vanishes with the cut size as expected, so this solution resolves the unphysical behaviour of the expression in \eqref{eq:cutr} computed from the leading-order solution. Remarkably, we find using \textsc{Mathematica} that $\varrho\rightarrow \smash{\sqrt{3h/2}}(1-k)$ as $\sigma\rightarrow\infty$, recovering the result for $\Sigma=O(1)$ obtained earlier.

This solution resolves the unphysical behaviour in \eqref{eq:cutd} for the cut displacement in a similar way: The leading-order equations for the radial and vertical displacements of the shell are now
\begin{align}
&\dfrac{\mathrm{d}u}{\mathrm{d}\xi}=-\delta^3\left[\dfrac{b^2}{2}+(\sigma+\xi)b-f\right],&&\dfrac{\mathrm{d}v}{\mathrm{d}\xi}=\delta^2 b,\label{eq:uv}
\end{align}
where $f(\xi)=[n(\xi)-e(\xi)]/2=[n(\xi)-(\sigma+\xi)\dot{n}(\xi)]/3$ using the first of \eqsref{eq:proba}. The first equation cannot be integrated in closed form, so the approximate solution does not yield a correction to the expression for the cut opening in a straightforward manner. However, \eqsref{eq:uv} show that $v\gg u$ for $\Sigma=O(\delta)$, whence the cut edge displacement is $d\sim|v(\xi=0)-v(\xi\rightarrow\infty)|$. This leads to
\begin{align} 
d=\dfrac{\varsigma\left[\operatorname{ker}_1{\varsigma}\left(\operatorname{kei}_0{\varsigma}-\operatorname{ker}_0{\varsigma}\right)-\operatorname{kei}_1{\varsigma}\left(\operatorname{kei}_0{\varsigma}+\operatorname{ker}_0{\varsigma}\right)\right]}{\sqrt{2}\left\{\left(\operatorname{ker}_1{\varsigma}\right)^2+\left(\operatorname{kei}_1{\varsigma}\right)^2+\sqrt{2}\varsigma\left[\operatorname{ker}_1{\varsigma}\left(\operatorname{ker}_0{\varsigma}+\operatorname{kei}_0{\varsigma}\right)+\operatorname{kei}_1{\varsigma}\left(\operatorname{kei}_0{\varsigma}-\operatorname{ker}_0{\varsigma}\right)\right]\right\}}h|1-k|.
\end{align} 
We find $d\rightarrow h|1-k|/2$ as $\sigma\rightarrow\infty$, which recovers the result for $\Sigma=O(1)$. This is quite remarkable, because the radial displacement, which is negligible for $\Sigma=O(\delta)$ contributes to the expression for $d$ obtained for $\Sigma=O(1)$. Moreover, $d\rightarrow 0$ as $\sigma\rightarrow 0$, resolving the limitation of the solution for $\Sigma=O(1)$ as announced.

\subsection*{Recoil velocity: scaling argument} 
With the leading-order asymptotic solution, we can understand the scaling of the velocity of the recoil from the undeformed to the deformed solution.
If the shell were to remain in its undeformed configuration,
\begin{align}
&E_s=E_\phi=0,&&K_s=K_\phi=1-k. 
\end{align}
The deformed solution has $N_s\ll E_\phi$ from \eqsref{eq:scalings}, whence, from the first of \eqsref{eq:stresses}, $E_s=-E_\phi/2+O(\delta)$. It follows that, in the asymptotic inner region,
\begin{subequations}
\begin{align}
&E_s\sim-\tfrac{1}{2}\delta^2e,&&E_\phi\sim\delta^2e,
\end{align}
so definitions~\neqref{eq:str} imply that $f_s,f_\phi\sim 1$. Hence, from \eqsref{eq:str} and \neqref{eq:scalings},
\begin{align}
&K_s\sim 1-k+\dot{b},&&K_\phi\sim 1-k. 
\end{align}
\end{subequations}
The elastic energy density~\cite{haas15} is thus
\begin{align}
E_s^2+E_sE_\phi+E_\phi^2+\varepsilon^2\left(K_s^2+K_sK_\phi+K_\phi^2\right)\sim\left\{\begin{array}{cl}
3\varepsilon^2(1-k)^2&\text{in the undeformed configuration},\\
\dfrac{3}{4}\delta^4e^2+\varepsilon^2\left[3(1-k)(1-k+\dot{b})+\dot{b}^2\right]&\text{in the deformed configuration}.
\end{array}\right. 
\end{align}
On noting that $e=\tfrac{4}{3}\dddot{b}$ from \eqsref{eq:diffeqs}, it follows that the elastic energy released during the recoil from the undeformed to the deformed configuration is
\begin{align}
\Delta\mathcal{E}&=2\pi\int_0^\infty{\delta^4\left[\dfrac{4}{3}\dddot{b}\,^2+\dot{b}^2+3\dot{b}(1-k)\right]\sin{\Sigma}\,(\delta\,\mathrm{d}\xi)}=-3^{7/4}\pi\delta^5(k-1)^2\sin{\Sigma}\leq 0,
\end{align}
using the leading-order solution \neqref{eq:bsol} and in which the factor $\sin{\Sigma}$ is the leading-order Jacobian of polar coordinates. Of course, the fact that elastic energy is released is hardly surprising, but we needed to compute this number to find the leading-order scaling of this energy release.

We now make this result dimensional: in our derivation of the leading-order solution, we have implicitly nondimensionalised distances with the dimensional radius $R$ of the shell, and energies with $YR^2$, where $Y=EH$ is the Young's modulus of the shell, with $E$ its elastic modulus and $H$ is dimensional thickness, so that $h=H/R$. On unravelling this nondimensionalisation, the dimensional energy released during the recoil thus scales as $\Delta\mathcal{E}\sim EHR^2\delta^5(k-1)^2\sin{\Sigma}$. During the recoil, the sheet moves a (dimensional) distance $D\sim R\delta^2|1-k|\sin{\Sigma}\sim H|1-k|\sin{\Sigma}$ from \eqref{eq:cutd}. The effective elastic force driving the recoil is thus $F\sim \Delta\mathcal{E}/D\sim EHR\delta^3|1-k|\sim EH^{5/2}R^{-1/2}|1-k|$. The recoil is resisted by viscous dissipation in the tissue and the surrounding fluid. The hydrodynamic strain rate scales as $\mu V/(\delta R)$, where $\mu$ is an effective viscosity and $V$ is the initial recoil velocity, i.e. the fluid velocity that decays over the extent $\delta R$ of the asymptotic inner region. The later has surface area $R^2\delta$, and so $F\sim (\mu V/\delta R)(R^2\delta)\sim \mu VR$. The initial recoil velocity and hence the hydrodynamic force are independent, at leading order, of $k$ because the initial shape of the cell sheet is, by the results of toy problem 1. The force balance of elastic and hydrodynamic forces then yields the scaling of the recoil velocity
\begin{align}
V\sim\dfrac{E}{\mu}\dfrac{H^{5/2}}{R^{3/2}}|1-k|.
\end{align}
This scaling breaks down for small cut sizes, because it does not satisfy $V\rightarrow 0$ as $\Sigma\rightarrow 0$, but resolving this would require a more detailed understanding of the dependence of the fluid velocity on cut size and its decay away from the rim of the cut, which is beyond the scope of our analysis.

\section*{Toy Problem 3: circular ablation of a spherical elastic shell with anisotropic contraction}
Finally, we show how (spatially uniform) anisotropic contraction of a spherical shell of unit radius leads to stresses (and hence to recoil on ablation) irrespective of the intrinsic curvatures of the shell. Let $f_1\not=f_2$ be the principal contractions, and take cylindrical coordinates such that the (orthogonal) principal directions of contraction align with the azimuthal and circumferential directions of the shell. As in the previous section, we denote by $s$ and $S(s)$ the undeformed and deformed arclengths of the shell along its meridian (measured from a pole), and by $\rho(s)=\sin{s}$ and $r(s)$ the undeformed and deformed polar radii of the shell. The strains of the midsurface of the shell~\cite{haas21} are thus
\begin{align}
&E_s(s)=\dfrac{f_s(s)}{f_1}-1,&& E_\phi(s)=\dfrac{f_\phi(s)}{f_2}-1,&&\quad\text{where }f_s(s)=S'(s),\;f_\phi(s)=\dfrac{r(s)}{\rho(s)},
\end{align}
and where dashes denote differentiation with respect to $s$. Let $\psi(s)$ denote the deformed tangent angle to the shell, which satisfies $r'(s)=f_s\sin{\psi}$. At the pole of the shell, we have $r(0)=\rho(0)=\psi(0)=0$ by geometric continuity, so l'Hôpital's rule implies $f_\phi(0)=r'(0)/\rho'(0)=f_s(0)\cos{\psi(0)}/\!\cos{0}=f_s(0)$. Since $f_1\not=f_2$, it follows that $E_s(0)$ and $E_\phi(0)$ cannot both vanish, so the meridional and circumferential stresses $N_s=2E_s+E_\phi$ and $N_\phi=E_s+2E_\phi$, the expressions of which are derived in Ref.~\cite{haas15}, cannot both vanish near the pole by continuity, and hence the shell will recoil on ablation.

While these stresses are caused by anisotropic contraction, we note that they also generate a curvature mismatch. Indeed, let $K_s,K_\phi$ denote the mismatch of the actual and intrinsic curvatures of the shell in the meridional and circumferential directions. Then $M_s=2K_s+K_\phi$, $M_\phi=K_s+2K_\phi$ satisfy the differential equation $\varepsilon^2(rM_s)'=f_s(rN_s\tan{\psi}+\varepsilon^2M_\phi\cos{\psi})$ as derived in Ref.~\cite{haas15}, where $\varepsilon$ is again the (scaled) relative thickness of the shell. Hence $M_s,M_\phi$ cannot both vanish identically since $N_s$ does not, and so neither can $K_s,K_\phi$ as claimed.

\section*{Fitting of the morphoelastic theory to the average cell sheet midlines of \emph{Volvox} inversion}
We fit our morphoelastic model (\emph{Materials and Methods} of the main text), based on the elastic shell theory for large bending deformations of Ref.~\cite{haas21}, to the average cell sheet midlines of \emph{Volvox} inversion~\cite{haas18a} similarly to the fitting described in Ref.~\cite{haas18a}. The use of the upgraded shell theory of Ref.~\cite{haas21} is the main mechanical difference between the quantitative inversion model of this paper and that of Ref.~\cite{haas18a}. In what follows, we describe the quantitative model and the fitting in detail.
\begin{figure}[b!]
\centering\includegraphics{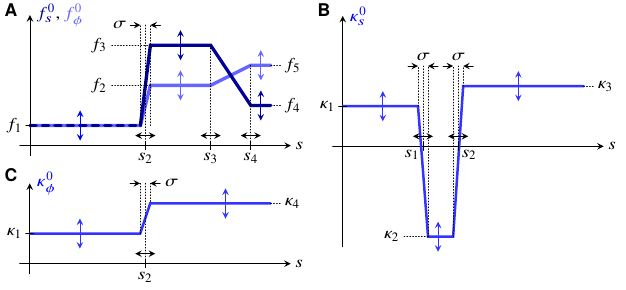} 
\caption{Functional forms of intrinsic stretches and curvatures. (A) Plots of the piecewise linear or constant functional forms of the intrinsic stretches $f_s^0,f_\phi^0$ against arclength~$s$ (\emph{Materials and Methods} of the main text), defining fitting parameters $f_1,f_2,f_3,f_4,f_5$, $s_2<s_3<s_4$ encoding the intrinsic stretches of the different cell shapes observed during \emph{Volvox} inversion~(S\citealp{haas18a},S\citealp{hohn11}) and the positions of the boundaries between regions of different cell shapes. (B) Plot of the meridional intrinsic curvature $\kappa_s^0$ against $s$, defining additional fitting parameters $\kappa_1,\kappa_2,\kappa_3$ encoding the intrinsic curvatures of the different cell shapes and the additional positional fitting parameter $s_1<s_2$. (C) Plot of the circumferential intrinsic curvature $\kappa_\phi^0$ against $s$, defining the additional fitting parameter $\kappa_4$ that defines yet another intrinsic curvature value for a cell shape. As in Ref.~\cite{haas18a}, the parameter $\sigma$ regularising what would otherwise be discontinuities in $\smash{f_s^0,f_\phi^0,\kappa_s^0,\kappa_\phi^0}$ is not fitted for.}\label{figS}
\end{figure}
\subsection*{Fitting parameters encode cell shape changes in intrinsic stretches and curvatures} Similarly to Ref.~\cite{haas18a}, we define piecewise constant or linear functional forms for the intrinsic stretch and intrinsic curvature functions $f_s^0,f_\phi^0,\kappa_s^0,\kappa_\phi^0$ (\emph{Materials and Methods} of the main text) that encode the cell shape changes observed during \emph{Volvox} inversion~\cite{hohn11}. These functional forms are shown in Fig.~\ref{figS}. They define fitting parameters $f_1,f_2,f_3,f_4,f_5$, $\kappa_1,\kappa_2,\kappa_3,\kappa_4$, $s_1<s_2<s_3<s_4$ that encode the intrinsic stretches and intrinsic curvatures of the different cell shapes observed during \emph{Volvox} inversion, and the positions of the boundaries between regions of different cell shapes. There are thus 13 fitting parameters. Since the intrinsic stretches and curvatures change during development, as the programme of cell shape driving inversion is run through, each of these parameters is \emph{a priori} a function of time, so has to be fitted for at each developmental timepoint. This number of fitting parameters may therefore appear to be large, but we emphasise, as discussed in detail in Ref.~\cite{haas18a}, that the functional forms of intrinsic stretches and curvatures are the minimal ones that can encode the cell shape changes observed during \emph{Volvox} inversion~\cite{hohn11} and are appropriately continuous. 
\subsection*{Fitting methods and details} For fitting this model to the average shape of \emph{Volvox} at a given developmental timepoint of inversion, similarly to Ref.~\cite{haas18a}, we place fitting points $(R_i,Z_i)$ for $i=1,2,\dots,N_{\text{fit}}=100$ at equal spacing along the arclength of the meridian of this average shape. For a given set of parameter values, we solve the boundary value problem associated with Eqs.~\textbf{7} of the main text using the \texttt{bvp4c} solver of \textsc{Matlab} (The MathWorks, Inc.) or a custom implementation of arclength continuation based on the same solver. The solution of this boundary value problem defines a cross-section $(r(s),z(s))$, where $\mathrm{d}z/\mathrm{d}s=f_s\sin{\psi}$ (\emph{Materials and Methods} of the main text). This yields points $(r_i,z_i)$ equally spaced along this cross-section, for $i=1,2,\dots,N_{\text{fit}}=100$. The fitting minimises the fit energy
\begin{align}
E_{\text{fit}}=\left\{\sum_{i=1}^{N_{\text{fit}}}{\bigl[(r_i-R_i)^2+(z_i-Z_i)^2\bigr]}\right\}^{1/2} 
\end{align}
over the fitting parameters. We perform this minimisation using a modification of the \texttt{fminsearch} function of \textsc{Matlab} that uses the modified parameters for its underlying Nelder--Mead algorithm suggested by Ref.~\cite{gao12} and that restricts the initial simplex used by the algorithm. The second modification avoids parameter values that are ``too far way'' from the initial guess and that might lead to the algorithm failing due to bifurcations that lie between the initial guess and these parameter values in parameter space. Further, we stopped the minimisation once $E_{\mathrm{fit}}$ dropped below a certain value (Fig. 3N of the main text); empirically, we found that this avoids excessive variations of fit parameters. We fit the average shapes for the developmental timepoints $t_1<t_2<\cdots$ one after the other, using the fitted parameter values from the previous timepoints to obtain initial guesses.
\subsubsection*{Fitting details for the different fits in the main text} For the fits discussed in the main text, the number of fitting parameters was reduced as follows: For the fits of constant $\kappa_1=\kappa_{\mathrm{post}}^0\in\{0,R^{-1},r_{\mathrm{post}}^{-1}\}$, the parameters that were fitted for at each timepoint were $f_2,f_3,f_4,f_5,\kappa_2,s_1,s_2,s_3,s_4$. Additionally, $f_1=r_{\mathrm{post}}/R,\kappa_3,\kappa_4$ were fitted only at the first timepoint and held constant for later timepoints. This assumption reduces the number of fitting parameters. For the final ``free'' fit, $f_1=r_{\mathrm{post}}/R,\kappa_1=\kappa_{\mathrm{post}}^0,\kappa_3,\kappa_4$ were additionally fitted for at each timepoint. 
\subsection*{Further discussion of the fitting results} 
\begin{figure}[h!]
\centering\includegraphics{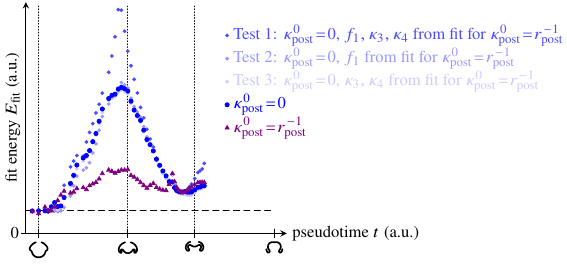} 
\caption{Test of the fit results in the main text. Plot of the fit energy $\smash{E_{\mathrm{fit}}}$ against developmental pseudotime $t$: Repeating the fits for $\smash{\kappa^0_{\mathrm{post}}}=0$ using the values of the parameters $f_1=r_{\mathrm{post}}$ and/or $\kappa_3,\kappa_4$ from the (better) fit with $\smash{\kappa_{\mathrm{post}}^0=r_{\smash{\mathrm{post}}}^{-1}}$ does not lead to better fits. The energies of the fits for $\smash{\kappa_{\mathrm{post}}^0\in\{0,r_{\smash{\mathrm{post}}}^{-1}\}}$ are repeated from Fig.~3N of the main text for comparison. Pseudotimepoints used for the discussion in Fig. 3D--L of the main text are highlighted by the shape insets on the horizontal axis.}\label{figS2}
\end{figure}

In this final subsection, we discuss the fitting results of the main text further. In particular, we must discuss the fact that we did not fit the parameters $f_1,\kappa_3,\kappa_4$ at each timepoint. Based on the observed cell shape changes~\cite{hohn11}, no variation of $f_1$ is expected while the cells in the posterior are spindle-shaped. This is borne out by the ``free'' fit, which suggests that $f_1$ remains (approximately) constant until around the time of phialopore opening, when the cells in the uninverted posterior become pencil-shaped~\cite{hohn11}, as discussed in the main text. Not fitting the parameters $\kappa_3,\kappa_4$ is \emph{a priori} a simplifying assumption, because the preferred curvatures of the teardop-shaped cells and disc-shaped cell shapes that are observed in the anterior hemisphere during early and late inversion~\cite{hohn11} could be different. Could the worse fit energies for the fits with $\smash{\kappa^0_{\mathrm{post}}\in\{0,R^{-1}\}}$ (Fig. 3N of the main text) result from the different values of $r_{\mathrm{post}}$ (inset of Fig. 3O in the main text) or $\kappa_3,\kappa_4$ fitted at the first timepoint? To exclude this possibility, we have run additional fits with $\smash{\kappa^0_{\mathrm{post}}}=0$ using the values of $f_1=r_{\mathrm{post}}$ and/or $\kappa_3,\kappa_4$ used for $\smash{\kappa_{\mathrm{post}}^0=r_{\mathrm{post}}^{-1}}$. These fits did not lead to better fit energies~(Fig.~\ref{figS2}). The fact that the variations of the fitted values of $\kappa_{\mathrm{post}}^0$ and $r_{\mathrm{post}}$ in the final ``free'' fit (Fig. 3O of the main text) are comparable to the differences between these values in the earlier fits is also consistent with this. All of this shows that these simplifying assumptions do not cause the worse fits, and hence that these result indeed from the ``wrong'' values of $\kappa_{\mathrm{post}}^0$.
\bibliography{supplement}